\documentclass{article}
\usepackage[T1]{fontenc}
\usepackage[utf8]{inputenc}

\usepackage[swedish,english]{babel} 

\usepackage{csquotes}

\usepackage{lscape}
\usepackage{amsmath}
\usepackage{mathtools}
\usepackage{amsfonts}
\usepackage{amssymb}
\usepackage{algorithm}
\usepackage{algpseudocode}
\usepackage{enumerate}
\usepackage{physics}
\usepackage{hyperref}
\usepackage{todonotes}
\usepackage{bbm}
\usepackage{longtable}
\usepackage{url}
\usepackage{color}
\usepackage{graphicx}
\usepackage{epstopdf}
\usepackage{multirow}
\usepackage{bm}

\usepackage{natbib}
\usepackage{verbatim}
\usepackage{float}
\usepackage[toc,page]{appendix}
\usepackage[onehalfspacing]{setspace}
\usepackage{adjustbox}
\usepackage{rotating}
\usepackage{amsthm}
\usepackage{booktabs,dcolumn,caption}
\usepackage{tikz}

\usepackage{arydshln}
\usepackage{array}

\usepackage{booktabs,caption}
\usepackage{subcaption}
\usepackage[affil-it]{authblk} 
\let\tilde\widetilde

\makeatletter
\newcommand*\rel@kern[1]{\kern#1\dimexpr\macc@kerna}
\newcommand*\widebar[1]{%
  \begingroup
  \def\mathaccent##1##2{%
    \rel@kern{0.8}%
    \overline{\rel@kern{-0.8}\macc@nucleus\rel@kern{0.2}}%
    \rel@kern{-0.2}%
  }%
  \macc@depth\@ne
  \let\math@bgroup\@empty \let\math@egroup\macc@set@skewchar
  \mathsurround\z@ \frozen@everymath{\mathgroup\macc@group\relax}%
  \macc@set@skewchar\relax
  \let\mathaccentV\macc@nested@a
  \macc@nested@a\relax111{#1}%
  \endgroup
}
\makeatother

\makeatletter
\def\FirstLetterUppercase#1{\expandafter\FirstLetterUppercase@i#1 \@nil}
\def\FirstLetterUppercase@i#1#2 #3\@nil{%
  \MakeUppercase{#1}#2
  \ifx\relax#3\relax\def\next@i{}\else\def\next@i{\expandafter\FirstLetterUppercase@i#3\@nil}\fi%
  \next@i}
\makeatother

\newcolumntype{L}[1]{>{\raggedright\let\newline\\\arraybackslash\hspace{0pt}}p{#1}}
\newcolumntype{C}[1]{>{\centering\let\newline\\\arraybackslash\hspace{0pt}}p{#1}}
\newcolumntype{R}[1]{>{\raggedleft\let\newline\\\arraybackslash\hspace{0pt}}p{#1}}

\newcommand{\argmax}{\operatornamewithlimits{arg\,max}}
\newcommand{\argmin}{\operatornamewithlimits{arg\,min}}

\title{DIF Analysis with Unknown Groups and Anchor Items}

\author[1,2]{Gabriel Wallin}
\author[1]{Yunxiao Chen}
\author[1]{Irini Moustaki}

\affil[1]{London School of Economics and Political Science}
\affil[2]{Ume{\aa} University}
 
\date{}

\begin{document}

\maketitle

\begin{abstract}
\noindent 

Ensuring fairness in instruments like survey questionnaires or educational tests is crucial. One way to address this is by a Differential Item Functioning (DIF) analysis, which examines if different subgroups respond differently to a particular item, controlling for their overall latent construct level. DIF analysis is typically conducted to assess measurement invariance at the item level. Traditional DIF analysis methods require knowing the comparison groups (reference and focal groups) and anchor items (a subset of DIF-free items). Such prior knowledge may not always be available, and psychometric methods have been proposed for DIF analysis when one piece of information is unknown. More specifically, when the comparison groups are unknown while anchor items are known,  latent DIF analysis methods have been proposed that estimate the unknown groups by latent classes. When anchor items are unknown while comparison groups are known,  methods have also been proposed, typically under a sparsity assumption -- the number of DIF items is not too large. However, DIF analysis when both pieces of information are unknown has not received much attention. This paper proposes a general statistical framework under this setting. In the proposed framework, we model the unknown groups by latent classes and introduce item-specific DIF parameters to capture the DIF effects. Assuming the number of DIF items is relatively small, an $L_1$-regularised estimator is proposed to simultaneously identify the latent classes and the DIF items. A computationally efficient Expectation-Maximisation (EM) algorithm is developed to solve the non-smooth optimisation problem for the regularised estimator. 
The performance of the proposed method is evaluated by simulation studies and an application to item response data from a real-world educational test.

\end{abstract}

\section{Introduction}
Psychometric models to analyse data from instruments such as survey questionnaires and educational tests rely on an equivalence assumption on the item parameters across groups of respondents. That is, conditioning on the latent construct measured by the instrument, a respondent's response to each item is independent of their group membership.
This assumption is known as measurement invariance. If violated, the psychometric property of the item(s) is not constant across groups, which can cause measurement bias \citep{millsap2012statistical}. The measurement invariance assumption is typically investigated through differential item functioning (DIF) analysis, a class of statistical methods that compares respondent groups at the item level and detects non-invariant, i.e., DIF, items.  

Traditional DIF detection methods assume that both the comparison groups and a set of non-DIF items, commonly referred to as the anchor set, are known a priori. The anchor items are used to identify the latent construct the instrument measures, and a DIF detection method compares the performances of the comparison groups, controlling for their performance on the anchor items as a proxy of the latent construct level.
Depending on their specific assumptions, these DIF detection methods can be divided into  Item-Response-Theory-based (IRT-based) methods \citep[e.g.,][]{lord1977study,lord1980applications,thissen1988data,wainer2012item, thissen2013use,kim1995detection,steenkamp1998assessing, tay2016item,woods2013langer} and non-IRT-based methods
 \citep[e.g.,][]{cao2017monte,holland1993differential,holland1986differential,dorans1986demonstrating,tay2015overview,swaminathan1990detecting,shealy1993model,woods2013langer,zwick2000using,drabinova2017detection}; see 
\citet{millsap2012statistical} for a review of traditional DIF analysis methods. Generally speaking, IRT-based methods tend to provide a clearer definition of DIF effects through a generative probabilistic model at the price of a risk of model misspecification.

Unfortunately, comparison groups and anchor items may not always be available in real-world applications, in which cases the aforementioned traditional methods are not applicable. Even if we have some information about anchor items, the result may be sensitive to the specific anchor items we use, when we only have a small number of such, and there will be a big issue if the anchor items are misspecified. Modern DIF analysis methods have been developed in situations where either the comparison groups or the anchor items are unknown. When anchor items are unknown, the latent construct is not identified, in which case, DIF detection is an ill-posed problem if no additional assumptions are made.  A reasonable assumption in this situation is sparsity -- the number of DIF items is relatively small, under which the detection of DIF items is turned into a model selection problem. To tackle the model selection problem, 
item purification methods have been proposed \citep[e.g.,][]{candell1988iterative,clauser1993effects,fidalgo2000effects,wang2003effects,wang2004effects,wang2009mimic,kopf2015anchor,kopf2015framework}, where stepwise model selection methods are used to detect DIF items. More recently, Lasso-type regularised estimation methods have been proposed to solve the model selection problem \citep{magis2015detection, tutz2015penalty, belzak2020improving, bauer2020simplifying, schauberger2020regularization}. In these methods, the DIF effects are represented by item-specific parameters under an IRT model, where a zero coefficient encodes no DIF effect for an item, and Lasso-type penalties are imposed on the DIF parameters to obtain a sparse solution, i.e., many items are DIF-free.  A drawback of regularised estimation methods is that, due to the bias brought by Lasso regularisation, they do not provide valid $p$-values for testing whether each item is DIF-free. Recently, 
\citet{chen2021dif} considered a limiting case of a regularised estimator and showed that the estimator can 
simultaneously identify the latent construct and yield valid statistical inferences on the individual DIF effects. 
An alternative direction of DIF analysis without anchor items is based on the idea of differential item pair functioning.
Under the Rasch model, \citet{bechger2015statistical} showed that 
although a Rasch model with group-specific difficulty parameters is not identifiable, the relative difficulties of item pairs are identifiable and can be used for detecting DIF items. Based on this idea, 
\citet{yuan2021differential} introduced visualisation methods for DIF detection. We lastly point out that there is related literature on $L_1$ regularisation for general mixture models such as Gaussian mixture models \citep[e.g.,][]{luo2008mixture,bhattacharya2014lasso,bouveyron2014model}, which also consider model-based clustering but are based on continuous instead of categorical data. 
However, it is worth noting that
these works all consider a high-dimensional data setting, and the regularisation is used for dimension reduction. The current paper focuses on a relatively low-dimensional setting, and a regularised estimator is proposed for the purpose of model selection.

The comparison groups may sometimes be unavailable, and DIF analysis in this situation is typically referred to as latent DIF analysis \citep{cho2016ncme,de2011explanatory}.
As suggested in \citet{de2011explanatory}, latent DIF analysis is needed when we do not know the crucial groups for comparison, we cannot observe the groups of interest, or there are validity  concerns regarding the true group membership of the respondents. For example, for self-reported health and mental health instruments \citep{teresi2016epilogue, reeve2016overview,teresi2021differential}, many covariates are collected, such as age, gender, ethnicity, and other background variables, but the crucial groups for DIF analysis are typically unclear. For another example, when analysing data from an educational test in which a subset of 
test takers have preknowledge on some leaked items \citep{cizek2017handbook}, the two comparison groups of interest -- the ones with and without item preknowledge -- are not directly observable.
Moreover, the observed group membership may sometimes poorly indicate the ``true'' group membership that causes the DIF pattern in the item response data 
\citep[e.g.,][]{bennink2014measuring, cho2010multilevel, finch2013investigation, von2011measuring}. Most existing latent DIF analysis methods assume that an anchor set is known and use a mixture IRT model -- a model that combines IRT and latent class analysis-- to identify the unknown groups and detect the DIF items simultaneously \citep{cho2010multilevel, cohen2005mixture, de2011explanatory}; see \citet{cho2016ncme}
for a review. 

In practice, both the comparison groups and the anchor set may be unknown. For example,  besides the aforementioned challenges of identifying the crucial comparison groups, the DIF analysis of self-reported health and mental health instruments also faces the challenge of identifying anchor items \citep{teresi2016epilogue, reeve2016overview}. In the item preknowledge example above, not only the comparison groups are unobserved, but also prior knowledge about non-leaked items is likely unavailable, and thus, correctly specifying an anchor set is a challenge \citep{o2016detecting}. Almost no general methods are available for latent DIF analysis when the anchor set is unavailable. Two notable exceptions are \citet{chen2022detection} and \citet{robitzsch2022regularized}. In \citet{chen2022detection}, a Bayesian hierarchical model for latent DIF analysis is proposed and applied this model to the simultaneous detection of item leakage and preknowledge in educational tests. 
In this model, latent classes are imposed among the test takers to model the comparison groups, and also among the items to model the DIF and non-DIF item sets. In addition, both the person- and item-specific parameters are treated as random variables and inferred via a fully Bayesian approach. However, the inference of this model relies on a Markov chain Monte Carlo algorithm, which suffers from slow mixing. Moreover, as most traditional DIF 
analysis methods adopt a frequentist setting, it is of interest to develop a frequentist approach to latent DIF analysis when the anchor set is unknown. \citet{robitzsch2022regularized} proposed a latent DIF procedure based on a regularised estimator under a mixture Rasch model. In this work, a nonconvex penalty called the  Smoothly Clipped Absolute Deviation (SCAD) penalty \citep{fan2001variable} other than the $L_1$ penalty is investigated. The methodology proposed in the current paper is similar in spirit to that of 
\citet{robitzsch2022regularized} but developed independently. The proposed framework focuses on the two-parameter logistic (2-PL) model \citep{birnbaum1968some} with an $L_1$ penalty and further provides a scope to generalise to other item response theory models. 

This paper proposes a frequentist framework for DIF analysis when both the comparison groups and the anchor set are unknown. The proposed framework combines the ideas of mixture IRT modeling for latent DIF analysis and regularised estimation for manifest DIF analysis with unknown anchor items. More specifically, the unknown groups are modelled by latent classes, and the DIF effects are characterised by item-specific DIF parameters. 
An $L_1$-regularised marginal maximum likelihood estimator is proposed, assuming that the number of DIF items is relatively small. This estimator penalises the DIF parameters by a Lasso regularisation term so that the DIF items can be selected by the non-zero pattern of the estimated DIF parameters. Computing the $L_1$-regularised estimator involves solving a non-smooth optimisation problem. We propose a computationally efficient Expectation-Maximisation (EM) algorithm \citep{dempster1977maximum, bock1981marginal}, where the non-smoothness of the objective function is handled by a proximal gradient method \citep{parikh2014proximal}. We evaluate the proposed method through simulation studies and an application to item response data from a real-world educational test. For the real-world application, we consider data from a midwestern university in the United States. This data set has been studied in \citet{bolt2002item}, where end-of-test items are believed to cause DIF due to insufficient time. Both the comparison groups, i.e. the speeded and non-speeded respondents, and the anchor items are unknown. In \citet{bolt2002item}, the DIF items and comparison groups are detected by borrowing information from an additional test form which is carefully designed so that the potential speededness-DIF items in the original form are administered at earlier locations, and thus, are unlikely to suffer from speededness-DIF. Thanks to the proposed procedure, we are able to identify the unknown DIF items and comparison groups without utilising information from the additional test form, and our findings are consistent with those of \citet{bolt2002item}. 

The rest of the paper is organised as follows. In Section \ref{sec:Model}, we propose a modelling framework for latent DIF analysis with unknown groups and anchor items and a regularised estimator that simultaneously identifies the unknown groups and detects the DIF items. In Section \ref{EMalgo}, we propose a computationally efficient EM algorithm. The proposed method is evaluated by 
simulation studies in Section \ref{Simulations} and further applied to data from a real-world educational test 
in Section \ref{EmpiricalAnalysis}. We conclude with discussions in Section~\ref{ConcludingRemarks}. Details about the computational algorithm are given in the Appendix.


\section{Proposed Framework}\label{sec:Model}


\subsection{Measurement Model} 
Consider $N$ respondents answering $J$ binary items. Let $Y_{ij} \in \{0, 1\}$ for $i = 1, \ldots, N$ and $j = 1, \ldots, J$ be a binary random variable recording individual $i$'s response to item $j$. The response vector of individual $i$ is denoted by $\mathbf{Y}_i = (Y_{i1}, \ldots, Y_{iJ})^\top$. We assume that the items measure a unidimensional construct, which is modelled by a latent variable $\theta_i$. We further assume that the respondents are random samples from $K+1$ unobserved groups, where the group membership is denoted by the latent variable $\xi_i \in \{0, 1, ..., K\}$.
Given the latent trait $\theta_i$ and the latent class $\xi_i$, consider the two-parameter item response model with a logit link (2-PL) (measurement model) \citep{birnbaum1968some}
\begin{equation}\label{eq:IRT_dif}
   \text{logit} P(Y_{ij} = 1|\theta_i, \xi_i ) = a_j \theta_i + d_j+  \delta_{j\xi_i},
\end{equation}
where $a_j$ and $d_j$ are known as the discrimination and easiness parameters respectively  and 
$\delta_{j \xi_i}$ is referred to as the DIF-effect parameter, as it 
quantifies the DIF effect of latent class $k$ on item $j$. 

We treat $\xi_i = 0$  as the baseline group, also known as the reference group, and set  $\delta_{j0} = 0$ for all $j = 1, \ldots, J$. In that case, $ a_j \theta_i + d_j$ denotes the item response function for the reference group. 
When $a_j$ is common across all items, the baseline model becomes the Rasch model \citep{rasch1960studies}. We focus on the 2-PL model here, but the proposed method easily adapts to other baseline IRT models.

The parameter $\delta_{jk}$ characterises how respondents in group $k$ differ from those in the reference group in terms of the item response behaviour on item $j$. For the reference group, the DIF parameter remains
zero for all items, serving as a reference point. For the remaining latent classes, the DIF parameter can be non-zero for certain items. Crucially, the magnitude of this parameter is allowed to differ across these latent classes. This flexibility accounts for varying degrees of DIF effects across different latent groups, when comparing with the reference group. The DIF effect parameter can also be expressed in terms of log-odds. Specifically, under the 2-PL model,  
$$\delta_{jk} = \log\left(\frac{P(Y_{ij} = 1|\theta_i = \theta, \xi_i = k)/(1-P(Y_{ij} = 1|\theta_i = \theta, \xi_i = k))}{P(Y_{ij} = 1|\theta_i = \theta, \xi_i = 0)/(1-P(Y_{ij} = 1|\theta_i = \theta, \xi_i = 0))}\right),$$
i.e., $\delta_{jk}$ is the log-odds-ratio when comparing two respondents from group $k$ and the reference group given that they have the same latent construct level.


\subsection{Structural Model}
The structural model specifies the joint distribution of the latent variables $(\theta_i, \xi_i)$. We assume that the latent classes follow a categorical distribution, 
$$
\xi_i \sim \textrm{Categorical}(\{0, 1, \ldots, K\}, (\nu_0, \nu_1, \ldots, \nu_K)),
$$
where $P(\xi_i = k) = \nu_k$. There are consequently $K+1$ latent classes with class probabilities $\boldsymbol\nu = (\nu_0, \nu_1, \ldots, \nu_K)^\top$ such that $\nu_k \geq 0$ and $\sum_{k=0}^{K} \nu_k = 1$. 
We further assume that conditional on $\xi_i$, the latent ability $\theta_i$ follows a normal distribution with class-specific mean and variance, i.e., 
$$
\theta_i | \xi_i = k \sim \mathcal{N}(\mu_k, \sigma_k^2).
$$ 
To ensure model identification, we fix the mean and variance of the reference group, i.e. $\mu_0 = 0$ and $\sigma_0^2 = 1$. 

\begin{figure} 
    \centering
    \includegraphics[width=0.8\textwidth, trim = 3cm 3cm 2cm 4cm, clip]{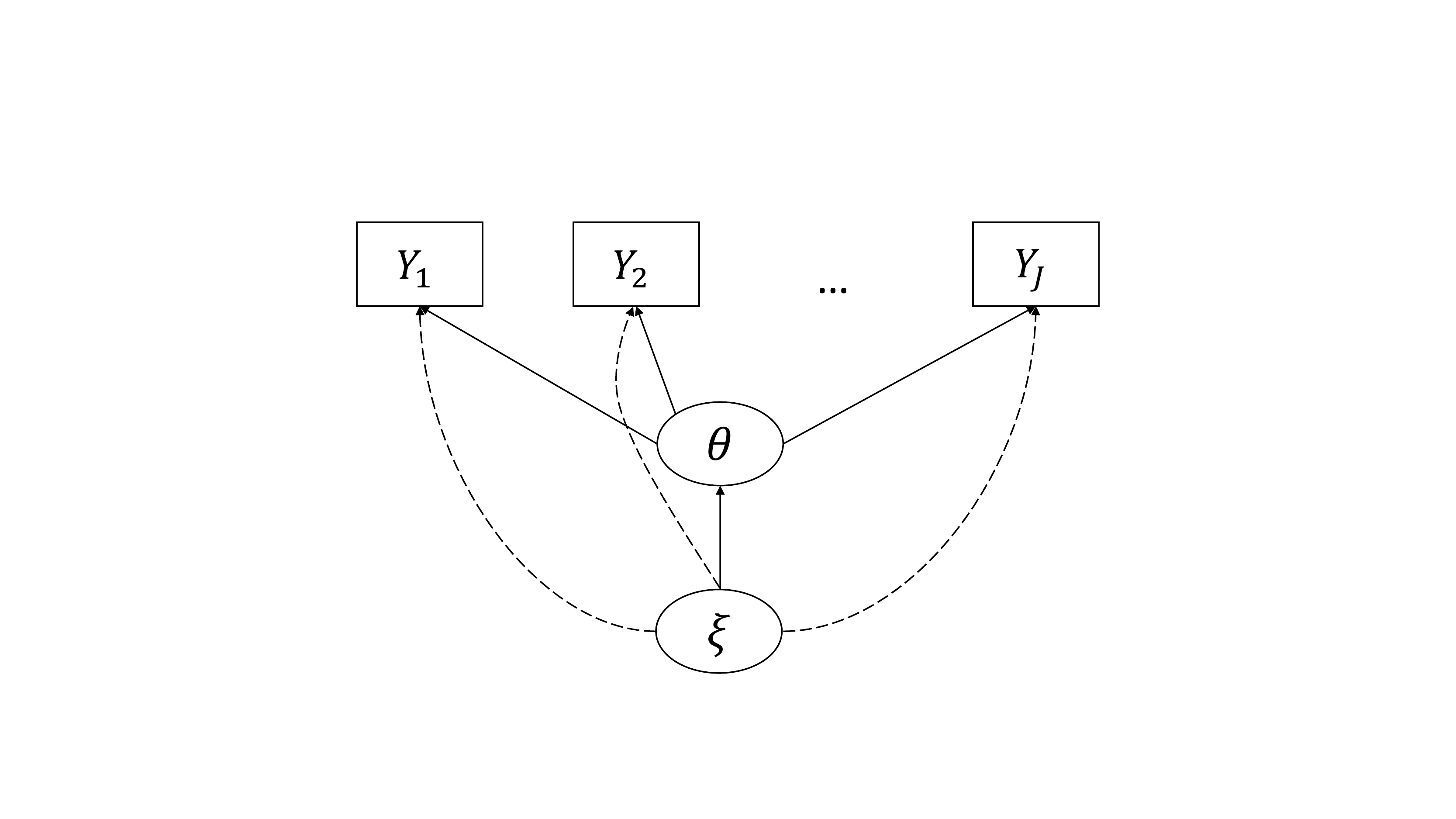}
    \caption{Path diagram of the proposed model, where the dashed lines indicate the DIF effects.}
    \label{fig:MIMIC}
\end{figure}

The path diagram of this model is given Figure~\ref{fig:MIMIC}. We note that the model coincides with a MIMIC model \citep{joreskog1975estimation} for manifest DIF analysis \citep{muthen1985multiple, muthen1989latent, woods2009evaluation, woods2011testing} when conditioning on the latent class $\xi_i$ (i.e., viewing $\xi_i$ as observed). However, since $\xi_i$ is unobserved, the statistical inference of the proposed model differs substantially from that of the MIMIC model. More specifically, the inference of the proposed model will be based on the 
marginal likelihood function where both latent variables $\xi_i$ and $\theta_i$ are marginalised out.  When the baseline IRT model is the 2-PL model, the marginal likelihood function takes the form
\begin{equation}\label{eq:llik}
\begin{split}
    L(\Delta) = \prod_{i=1}^N \sum_{k=0}^K \nu_k    \int \left(\prod_{j=1}^J \left(\exp((a_j\theta + d_j + \delta_{jk})Y_{ij})/(1+\exp(a_j\theta + d_j + \delta_{jk}))\right)\right) \phi(\theta \vert \mu_k, \sigma^2_k) d\theta,
\end{split}
\end{equation}
where $\phi(\theta \vert \mu_k, \sigma^2_k)$ denotes the density function of a normal distribution with mean $\mu_k$ and variance $\sigma_k^2$, and we use vector $\Delta$ to denote all the unknown parameters, including the item parameters  $a_j$ and $d_j$, $\delta_{jk}$, $\nu_k$, $\mu_k$ and $\sigma_k^2$, for $j = 1, ..., J$ and $k = 0, 1, ..., K$.

\subsection{Model Identifiability}

The current model suffers from two sources of unidentifiability. The first source of unidentifiability comes from not knowing the anchoring items, which occurs even if we condition on the latent class $\xi_i$, i.e., when the model becomes a MIMIC model for manifest DIF analysis. That is, for any constants $c_1, ..., c_K$, if we simultaneously replace $\mu_k$ by $\mu_k + c_k$ and replace $\delta_{jk}$ by $\delta_{jk} - a_j \mu_k$ for all $j = 1, ..., J$ and $k=1, ..., K$, the likelihood function value $L(\Delta)$ remains the same. This source of unidentifiability can be avoided when one or more anchor items are known a priori. Suppose that item $j$ is known to be DIF-free. Under the proposed model framework, it implies the constraints $\delta_{jk} = 0$ for all $k$.  Consequently, the aforementioned transformation can no longer apply, as otherwise, the zero constraints for the anchor items will be violated. As discussed in Section~\ref{subsec:reg}, this source of unidentifiability can be handled by a regularised estimation approach under a sparsity assumption that many DIF parameters $\delta_{jk}$ are zero. 

The second source of unidentifiability is the label-switching phenomenon of latent class models \citep{redner1984mixture}, as a result of the exchangeability of the latent classes. Under the current model, the baseline class is uniquely identified through the constraints $\delta_{j0} =0$, $\mu_0 = 0$ and $\sigma_0^2 = 1$. However,  the remaining latent classes are exchangeable, and the likelihood function value remains the same
when switching their labels. While label switching often causes trouble when inferring a  latent class model with Bayesian Markov chain Monte Carlo (MCMC) algorithms \citep{stephens2000dealing}, 
it is not a problem for the estimator to be discussed in Section~\ref{subsec:reg}. Our estimator is proposed 
under the frequentist setting and computed by an EM algorithm. 
When the EM algorithm converges, it will reach one of the equivalent solutions in the sense of label switching. 

\subsection{Sparsity, Model Selection and Estimation}\label{subsec:reg}

As explained above, the latent trait cannot be identified without anchor items. In that case, additional assumptions are needed to solve the DIF analysis problem. 
Specifically, we adopt the sparsity assumption, i.e., many DIF parameters $\delta_{jk}$ are zero. This is a common assumption in the manifest DIF literature, see for example \citet{magis2015detection, tutz2015penalty, belzak2020improving, bauer2020simplifying, schauberger2020regularization}. In many applications, for example, the detection of aberrant behaviour or parameter drift in educational testing, the number of DIF items is low, suggesting that this assumption is meaningful.  

Under the above sparsity assumption, we propose an $L_1$ regularised estimator to simultaneously estimate the unknown model parameters and learn the sparsity pattern of the DIF-effect parameters. This estimator takes the form 
\begin{equation}\label{criterion}
 \tilde{\Delta}^{(\lambda)} = \argmin_{\Delta}  -\log L(\Delta) + \lambda \sum_{j=1}^J\sum_{k=1}^K |\delta_{jk}|, \mbox{~s.t.~} \nu_k \geq 0, k=0, 1, ..., K, \mbox{~and~} \sum_{k=0}^K \nu_k =1, 
\end{equation}
where $L(\Delta)$ is the marginal likelihood function defined in \eqref{eq:llik}, and $\lambda > 0$ is a tuning parameter. The computation of 
this estimator will be discussed in Section~\ref{EMalgo}. 
Similar to  Lasso regression \citep{tibshirani1996regression}, the $L_1$ regularisation term $\lambda \sum_{j=1}^J \sum_{k=1}^K|\delta_{jk}|$ in \eqref{criterion} tends to shrink some of the DIF-effect parameters to be exactly zero. In the most extreme case where $\lambda$ goes to infinity, all the DIF-effect parameters will shrink to zero.  Under suitable regularity conditions and when $\lambda$ is chosen properly (i.e., $\lambda$ goes to infinity at a suitable speed), the $L_1$ regularised estimator yields both estimation and selection consistency \citep{zhao2006model,van2008high}. In that case, the latent trait is consistently identified, and the consistently selected sparse patterns of the estimated DIF-effect parameters can be used to classify items as DIF and non-DIF items.  

We select the tuning parameter $\lambda$ based on the Bayesian Information Criterion \citep[BIC;][]{schwarz1978estimating} using a grid search approach. Specifically, we consider a pre-specified set of grid points for $\lambda$, denoted by \\ $\lambda_1$, ... $\lambda_{M}$. For each value of $\lambda_m$, we solve the optimisation problem \eqref{criterion} and obtain $\tilde{\Delta}^{(\lambda_m)}$. To compute the BIC value for the model encoded by $\tilde{\Delta}^{(\lambda_m)}$, we compute a constrained maximum likelihood estimator, fixing $\delta_{jk}$  to zero if $\tilde{\delta}_{jk}^{(\lambda_{m})} = 0$. That is, 
\begin{equation}\label{eq:constrained}
\begin{aligned}
\hat{\Delta}^{(\lambda_m)} =& \argmin_{\Delta}  -\log L(\Delta), \\
\mbox{s.t.~}& \nu_k \geq 0, k =0, ..., K, ~ \sum_{k=0}^K \nu_k =1, \\
&\delta_{jk} = 0 \mbox{~~if~~} \tilde{\delta}_{jk}^{(\lambda_{m})} = 0, j = 1, ..., J, k=1, ..., K.
\end{aligned}
\end{equation}
The BIC corresponding to $\lambda_m$ is calculated as 
\begin{equation}\label{eq:BIC}
\mbox{BIC}_{\lambda_m} = -2\log L(\hat{\Delta}^{(\lambda_m)}) + \log(N)\mbox{Card}(\hat{\Delta}^{(\lambda_m)}),    
\end{equation}
where $\mbox{Card}(\hat{\Delta}^{(\lambda_m)})$ denotes the number of free parameters in $\hat{\Delta}^{(\lambda_m)}$ that equals to the total number of free parameters in $\Delta$ minus the corresponding number of zero constraints. The tuning parameter is then selected as 
$$\hat \lambda = \argmin_{\lambda_m, m=1, ..., M} \mbox{BIC}_{\lambda_m}.$$
Thanks to the asymptotic properties of the BIC \citep{shao1997asymptotic}, the true model will be consistently selected if it can be found by one of the tuning parameters. 

We use the constrained maximum likelihood estimator $\hat{\Delta}^{(\hat \lambda)}$ as the final estimator of the selected model and declare an item $j$ to be a DIF item if 
$\Vert (\hat \delta_{j1}^{(\hat \lambda)}, ..., \hat \delta_{jK}^{(\hat \lambda)})^\top\Vert \neq 0$. We summarise this procedure in Algorithm \ref{alg:BIC} below.


\begin{algorithm}
\caption{Regularised estimation and model selection.}\label{alg:BIC}
\textbf{Input:} Observed data $Y_{ij}$, $i=1, ..., N$ and $j = 1, ..., J$, and a grid of tuning parameters $ 0 < \lambda_1 < \ldots <   \lambda_M$. 

\medskip
For $m = 1, \ldots, M$,
\begin{enumerate}
\item Solve the optimisation problem \eqref{criterion} with tuning parameter 
$\lambda_m$ and obtain solution  $\tilde{\Delta}^{(\lambda_m)}$. 

\item Given $\tilde{\Delta}^{(\lambda_m)}$, solve the constrained optimisation problem \eqref{eq:constrained} and obtain $\hat{\Delta}^{(\lambda_m)}$. 

\item Calculate the BIC value $\mbox{BIC}_{\lambda_m}$ based on equation \eqref{eq:BIC} for the model encoded by $\tilde{\Delta}^{(\lambda_m)}$. 

\end{enumerate}

Select $\hat \lambda = \argmin_{\lambda_m, m=1, ..., M} \mbox{BIC}_{\lambda_m}.$ 
\medskip

\textbf{Output:} Estimate of the selected model, $\hat{\Delta}^{(\hat \lambda)}$, and the selected DIF items $\{j: \Vert (\hat \delta_{j1}^{(\hat \lambda)}, ..., \hat \delta_{jK}^{(\hat \lambda)})^\top\Vert \neq 0\}$. 
\end{algorithm}

\subsection{Other Inference Problems}

The latent class membership can be inferred by an empirical Bayes procedure, i.e., by the maximum a posteriori (MAP) estimate under the estimated model. For the MAP estimator, the goal is to find the most probable latent class $k$ for each respondent $i$, given their observed responses $\boldsymbol{y}_i$. This is done by maximizing the posterior probability of $\xi_i$ being equal to $k$, conditioned on the observed responses $\boldsymbol{y}_i$:
$$
\hat{\xi}_{\text{MAP}, i} = \argmax_{k \in \{0, 1, ..., K\}} \hat{P}(\Hat{\xi} = k|\mathbf{y}_i)
$$
We get these posterior probabilities through

\begin{equation}\label{eq:postprob}
\hat{P}(\xi_i = k | \mathbf{y}_i) = \frac{\hat{P}(\mathbf{y}_i | \xi_i = k, \theta_i) \cdot \hat{P}(\xi_i = k)}{\hat{P}(\mathbf{y}_i)}
\end{equation}
based on the estimated model parameters. 

Lastly, the number of latent classes, i.e., the choice of $K$, can be determined using the BIC. That is, we solve the optimisation problem in \eqref{criterion} for different values of $K$, yielding $\hat{\Delta}^{(K)}$. We thereafter compute the BIC as 
\begin{equation}\label{eq:BIC_corrected}
\mbox{BIC}(K) = -2\log L(\hat{\Delta}^{(K)}) + \log(N)\mbox{Card}(\hat{\Delta}^{(K)}),
\end{equation}
and select the $K$ that yields the smallest BIC:
\begin{equation}
    K^* = \argmin_K \mbox{BIC}(K)
\end{equation}

\subsection{Extensions}\label{sec:Extensions}

The proposed model is possible to extend in several ways to accommodate different data types and more than one factor. To make this clear, our model can be expressed as 
\begin{equation}\label{eq:IFA}
P(Y_{ij} = 1 | \theta_i, \xi_i) = f\bigg(g(\theta_i, \beta_j) + \delta_{j \xi_i} \bigg)
\end{equation}
The function $g$ determines the parametrisation of the person and item parameters, denoted by $\theta$ and $\beta$ respectively. If for example the Rasch model \citep{rasch1960studies} is adopted, $\beta_j = (a, d_j)$ where $a_j=a$ for all $j=1, \ldots, J$,
\begin{equation*}
g(\theta_i, \beta_j) = a\theta_i + d_j,
\end{equation*}
and $f(x) = \exp(x)/(1+\exp(x))$. 

It is also possible to consider link functions other than the logistic function considered in this paper, such as the probit link: 
$$
f(x) = \int_{- \infty}^x \phi(z) dz.
$$
DIF analysis using multidimensional IRT models with unknown anchor items has recently been considered in \citet{wang2021using}. As in the unidimensional case, no method can handle situations where both the groups and the anchor items are missing. Our proposed framework can however be extended to handle such situations. Consider the extension of \eqref{eq:IFA} where each respondent, in addition to the latent class membership $\xi_i$, is represented by an $L$-dimensional latent vector $\boldsymbol{\theta}_i = (\theta_{i1}, \ldots, \theta_{iL})^\top$. Each item is represented by an intercept parameter $d_j$ and $L$ loading parameters $ \boldsymbol{a}_j = (a_{j1}, \ldots, a_{jL})^\top$. This extension of the model can be expressed as a multidimensional 2-PL model with an added DIF component, i.e.,
\begin{equation*}
    P(Y_{ij} = 1 | \boldsymbol{\theta}_i, \xi_i) =  f\bigg(d_j + \boldsymbol{a}_j^\top \boldsymbol{\theta}_i + \delta_{j \xi_i} \bigg).   
\end{equation*}

The proposed modeling framework can also be extended to accommodate ordinal data. Denoting the response categories of $Y_{ij}$ by $m = 1, \ldots, M$, such model can, using the logistic link, be expressed as  
$$
\text{logit} P(Y_{ij} \leq m) = d_{jm} - a_j \theta_i + \delta_{j \xi_i}
$$
The model without the DIF parameter is known as the proportional odds model \citep{samejima1969estimation}. Note the negative sign in front of the slope parameter so that if $a_j$ is positive, increasing $\theta_i$ will increase the probability of higher-numbered levels of $Y_{ij}$. 

Lastly, we mention the possibility of extending the model to accommodate for DIF effects in the discrimination parameter, known as non-uniform DIF. To consider such a case, we introduce a similar DIF-effect parameter for $a_j$, just as we have for $d_j$. Let's denote the DIF effect on the discrimination parameter $\alpha_{j \xi_i}$. Given that $\xi_i=0$ is treated as the reference group, we set $\alpha_{j0}=0$ for all $j=1,\ldots, J$.  In this case, the modified 2-PL model with a logit link which accounts for DIF in both $a_j$ and $d_j$
can be written as:
\begin{equation}\label{eq:IRT_dif_modified}
\text{logit} P(Y_{ij} = 1|\theta_i, \xi_i ) = (a_j + \alpha_{j\xi_i}) \theta_i + d_j + \delta_{j\xi_i}.
\end{equation}

To include DIF in the $L_1$ regularised estimator, the penalty term needs to be modified to penalize both $\alpha_{jk}$ and $\delta_{jk}$ terms. The modified estimator is given by
\begin{equation}\label{criterion_modified}
\tilde{\Delta}^{(\lambda)} = \argmin_{\Delta} -\log L(\Delta) + \lambda \left( \sum_{j=1}^J\sum_{k=1}^K |\delta_{jk}| + \sum_{j=1}^J\sum_{k=1}^K |\alpha_{jk}| \right), \mbox{s.t.} \,\, \nu_k \geq 0, k=0, 1, ..., K, \mbox{and} \sum_{k=0}^K \nu_k =1,
\end{equation}


\section{Computation}\label{EMalgo}

The computation of the optimisation problems \eqref{criterion} and \eqref{eq:constrained} 
is carried out using the  EM algorithm \citep{dempster1977maximum, bock1981marginal}. An EM algorithm is an iterative algorithm, alternating between an Expectation (E) step and a Maximisation (M) step. 
Optimisation problem \eqref{eq:constrained} 
involves maximising the marginal likelihood function of a regular latent variable model, and
thus, can be solved by a standard EM algorithm. Thus, its details are skipped here. However, the optimisation problem \eqref{criterion} involves a non-smooth $L_1$ term. Consequently, the M step of the algorithm cannot be carried out using a gradient-based numerical solver, such as a Newton–Raphson algorithm. We develop an efficient proximal-gradient-based EM algorithm that uses a proximal gradient update \citep{parikh2014proximal} to carry out the non-smooth optimisation problem in the M-step. In what follows, we elaborate on this algorithm using the 2-PL model as the baseline IRT model, while pointing out that the algorithm easily extends to other baseline IRT models. 

Suppose that $t$ iterations of the algorithm have been run and let $\Delta^{(t)}$ be the current parameter value. In the E-step of the $t$th iteration, we construct a local approximation of the negative objective function at $\Delta^{(t)}$ in the form of 
\begin{equation} 
        Q(\Delta | \Delta^{(t)})  =  \sum_{i=1}^N \mathbb{E}\left[ \log\left(\nu_{\xi_i}\prod_{j=1}^J \left(\frac{\exp((a_j\theta_i + d_j + \delta_{j\xi_i})Y_{ij})}{1+\exp(a_j\theta_i + d_j + \delta_{j\xi_i})}\right) \phi(\theta_i \vert \mu_{\xi_i}, \sigma^2_{\xi_i}) \big\vert \mathbf Y_i, \Delta^{(t)}\right)\right]  - \lambda \sum_{j=1}^J\sum_{k=1}^K |\delta_{jk}|.
     \label{eq:E-step}
\end{equation}
We note that the expectation in \eqref{eq:E-step} is with respect to the conditional distribution of the latent variables $(\theta_i, \xi_i)$ given $\mathbf Y_i$, evaluated at the current parameters $\Delta^{(t)}$. 

In the M-step,  we find $\Delta^{(t+1)}$ such that 
$$
Q(\Delta^{(t+1)} | \Delta^{(t)}) > Q(\Delta | \Delta^{(t)}),
$$
or equivalently, 
$$-Q(\Delta^{(t+1)} | \Delta^{(t)}) < -Q(\Delta | \Delta^{(t)}).$$ 
By Jensen's inequality,  it consequently guarantees that the objective function of \eqref{criterion} decreases, i.e., 
$$
-\log L(\Delta^{(t+1)}) + \lambda \sum_{j=1}^J\sum_{k=1}^K |\delta_{jk}^{(t+1)}| < -\log L(\Delta^{(t)}) + \lambda \sum_{j=1}^J\sum_{k=1}^K |\delta_{jk}^{(t)}|.
$$
More specifically, we write 
$
\Delta = (\Delta_1^\top, \Delta_2^\top)^\top,
$ 
where $\Delta_1 = (\nu_0, ..., \nu_K)^\top$ and 
$\Delta_2$ contains the rest of the parameters. We notice that $-Q(\Delta | \Delta^{(t)})$ in \eqref{eq:E-step} can be decomposed as the sum of a smooth function \\
$
D_t(\Delta_1) = -\sum_{i=1}^N \mathbb{E}\left[ \log\left(\nu_{\xi_i} \big\vert \mathbf Y_i, \Delta^{(t)}\right)\right],
$
a smooth function $F_t(\Delta_2)$, defined as
$$
F_t(\Delta_2) = -\sum_{i=1}^N \mathbb{E}\left[ \log\left(\prod_{j=1}^J \left(\frac{\exp((a_j\theta_i + d_j + \delta_{j\xi_i})Y_{ij})}{1+\exp(a_j\theta_i + d_j + \delta_{j\xi_i})}\right) \phi(\theta_i \vert \mu_{\xi_i}, \sigma^2_{\xi_i}) \big\vert \mathbf Y_i, \Delta^{(t)}_2\right)\right]
$$ 
and a non-smooth function 
$
G(\Delta_2) =  \lambda \sum_{j=1}^J\sum_{k=1}^K |\delta_{jk}|.
$ 
We note that $\Delta_1^{(t+1)}$ can be obtained by solving the following constrained optimisation problem 
$$\Delta_1^{(t+1)} = \argmin_{\Delta_1} D_t(\Delta_1), \mbox{~s.t.~} \nu_k \geq 0, k=0, 1, ..., K, \mbox{~and~} \sum_{k=0}^K \nu_k =1.$$
Using the method of Lagrangian multiplier, this optimisation problem has a closed-form solution; see the Appendix for the details. 

We then find $\Delta_2^{(t+1)}$ such that $F_t(\Delta_2^{(t+1)}) + G(\Delta_2^{(t+1)}) < F_t(\Delta_2^{(t)}) + G(\Delta_2^{(t)})$. 
Consider the optimisation problem 
$\min F_t(\Delta_2) + G(\Delta_2)$. Due to the non-smoothness of $G$, this objective function is not differentiable everywhere. Consequently, gradient-based methods are not applicable. We find $\Delta_2^{(t+1)}$ by  using a proximal gradient method \citep{parikh2014proximal}. Denote the dimension of $\Delta_2$ by $d$, where $d $ equals the number of free parameters which is determined by the choice of baseline model ($2J + 2K$ for the 2-PL model). We define the proximal operator of $G$
at $\Delta_2$  as 
$$
\mbox{Prox}_G(\Delta_2) = \argmin_{\tilde\Delta_2 \in \mathbb R^d} ~G(\Delta_2) + \frac{1}{2} \Vert\tilde\Delta_2 - \Delta_2\Vert^2.
$$
We update $\Delta_2$ by 
\begin{equation}\label{eq:proximal}
\Delta_2^{(t+1)} = \mbox{Prox}_{\alpha G}(\Delta_2^{(t)} - \alpha \nabla F_t(\Delta_2^{(t)})),
\end{equation}
where $\nabla F_t(\Delta_2^{(t)})$ denotes the gradient of $F_t$ at $\Delta_2^{(t)}$, and $\alpha$ is a step size. According to \citet[][sect. 4.2]{parikh2014proximal}, for a sufficiently small step size $\alpha$, it is guaranteed that 
$F_t(\Delta_2^{(t+1)}) + G(\Delta_2^{(t+1)}) < F_t(\Delta_2^{(t)}) + G(\Delta_2^{(t)})$, unless 
$\Delta_2^{(t)}$ is already a stationary point. Thus, we select $\alpha$ by a line search procedure, whose details are provided in the Appendix. We note that this proximal operator has a closed-form solution. Specifically, each DIF-effect parameter $\delta_{jk}$ is updated by solving 
$$
\delta_{jk}^{(t+1)} = \argmin_{\tilde \delta_{jk}} -\frac{1}{2}\left(\tilde \delta_{jk} - \left(\delta_{jk}^{(t)} - \alpha \frac{\partial F_t(\Delta_2)}{\partial \delta_{jk}} \big\vert_{\Delta_2 = \Delta_2^{(t)}}\right)\right)^2 + \alpha\lambda |\tilde \delta_{jk}|,
$$
which has  a closed-form solution given by soft-thresholding \citep[Chapter 3,][]{hastie2009elements}. The rest of the parameters in $\Delta_2$ do not appear in the non-smooth function $G(\Delta_2)$, and thus, 
the resulting update of \eqref{eq:proximal} degenerates to a vanilla gradient descent update. For example, 
$$
d_j^{(t+1)} = d_j^{(t)}  - \alpha \frac{\partial F_t(\Delta_2)}{\partial d_{j}} \big\vert_{\Delta_2 = \Delta_2^{(t)}}.
$$
Further details of the proximal gradient update can be found in the Appendix. We summarise the main steps of this EM algorithm in Algorithm~\ref{alg:EM} below.

\begin{algorithm} 
\caption{An EM algorithm for solving \eqref{criterion}.}\label{alg:EM}
\textbf{Input:} The initial value $\Delta^{(0)} = (d_j^{(0)}, a_j^{(0)}, \delta_{jk}^{(0)}, \mu_k^{(0)}, \sigma_k^{(0)}, \nu_k^{(0)})$, $j=1, \ldots J, k = 0, \ldots, K$. 

\medskip

For iterations $t = 0, 1, 2, ..., T$, we iterate between the following steps: 

\begin{enumerate}
\item[] \textbf{E-Step:}  Calculate the $Q$ function in \eqref{eq:E-step}.

\item[] \textbf{M-step:} Calculate the parameter update for $\boldsymbol{\nu}$ according to
$$
\nu_k^{(t+1)}  = \frac{1}{N} \sum_{i=1}^{N} \gamma_{ik}^{(t)}, \,\,\,\,\, k = 1, \ldots K,
$$
where
$$
 \gamma_{ik}^{(t)} = \frac{\nu_k^{(t)}\prod_{j=1}^{J} \pi_{ijk}^{y_{ij}}(1-\pi_{ijk}^{1-y_{ij}}) \phi(\theta_i | \mu_k, \sigma_k^2) }{\sum_{k'=0}^K \nu_{k'}^{(t)} \prod_{j=1}^{J} \pi_{ijk}^{y_{ij}}(1-\pi_{ijk}^{1-y_{ij}}) \phi(\theta_i | \mu_k, \sigma_k^2)}
$$
and 
$$
\pi_{ijk} = P(Y_{ij} = 1 | \theta_i = \theta, \xi_i = k).
$$
Update the rest of the parameters according to
\begin{equation*}
    d_j^{(t+1)} = d_j^{(t)}  - \alpha \frac{\partial F_t(\Delta_2)}{\partial d_{j}} \big\vert_{\Delta_2 = \Delta_2^{(t)}}, \,\,\,\,\, j = 1, \ldots, J,
\end{equation*}
\begin{equation*}
    a_j^{(t+1)} = a_j^{(t)}  - \alpha \frac{\partial F_t(\Delta_2)}{\partial a_{j}} \big\vert_{\Delta_2 = \Delta_2^{(t)}}, \,\,\,\,\, j = 1, \ldots, J,
\end{equation*}
\begin{equation*}
    \mu_k^{(t+1)} = \mu_k^{(t)}  - \alpha \frac{\partial F_t(\Delta_2)}{\partial \mu_{k}} \big\vert_{\Delta_2 = \Delta_2^{(t)}}, \,\,\,\,\, j = 1, \ldots, J,
\end{equation*}
\begin{equation*}
    \sigma_k^{(t+1)} = \sigma_k^{(t)}  - \alpha \frac{\partial F_t(\Delta_2)}{\partial \sigma_{k}} \big\vert_{\Delta_2 = \Delta_2^{(t)}}, \,\,\,\,\, j = 1, \ldots, J,
\end{equation*}
and
\begin{equation*}
    \delta_{jk}^{(t+1)}=
    \begin{cases}
      \delta_{jk}^{(t)} - \alpha \lambda, & \text{if}\ 
      \delta_{jk}^{(t)} > \alpha \lambda \\
      \delta_{jk}^{(t)} + \alpha \lambda, & \text{if}\ 
      \delta_{jk}^{(t)} < - \alpha \lambda, \\
      0, & \text{otherwise}
    \end{cases}
  \end{equation*} 
where $\alpha$ is a step size chosen by line search. 

\end{enumerate}

Iterate until convergence has been reached.

\medskip
\textbf{Output:} $\Delta^{(T)} = (d_j^{(T)}, a_j^{(T)}, \delta_{jk}^{(T)}, \mu_k^{(T)}, \sigma_k^{(T)}, \nu_k^{(T)})$, $j=1, \ldots J, k = 0, \ldots, K$. 
\end{algorithm}
\vspace{5.5pt}
\noindent \textit{Remark. While differentiable approximations, such as the smoothed Lasso, can allow the use of gradient-based methods, they often come with their own set of challenges. Introducing a smoothing parameter can make the method sensitive to its choice, and in some situations, the approximation might not be close enough to the original problem, especially when the smoothing parameter is not sufficiently small. We opt for an approach based on non-smooth optimisation. We believe that directly tackling the non-smoothness ensures that we do not compromise on the sparsity of the solution, which is critical for our analysis. We have designed our algorithm around the EM algorithm, which inherently possesses certain convergence properties. Specifically, the EM algorithm is guaranteed to increase the log-likelihood in each iteration, and under mild regularity conditions, it converges to at least a local maximum of the log-likelihood. While we acknowledge that there are potential pitfalls in the convergence of non-smooth optimisation algorithms, we have implemented strategies to ensure the stability of our algorithm, such as adaptive step sizes and convergence checks. }

\section{Simulation Study}
\label{Simulations}
In this simulation, we evaluate the performance of the proposed method, treating the number of latent classes $K$ as fixed. We consider cases with two and three latent classes, respectively. For each simulation scenario, we run $B = 100$ independent replications.

\subsection{Settings}

We examine 16 scenarios under the two-group setting and 8 scenarios under the three-group setting, considering $J \in \{25, 50\}$, $N \in \{500, 1000\}$, and $K \in \{1, 2\}$. Note that $K=1$ represents a two-group setting and $K=2$ indicates a three-group setting. In the two-group setting, the class proportions are varied, with the reference group consisting of either 50\%, 80\%, or 90\% of the respondents. In the three-group scenario, 50\% of the respondents belong to the reference group, 30\% belong to the second latent class, and 20\% belong to the third latent class. 
The number of DIF items is set to 10 for all cases with $J=25$ and 20 for all cases with $J=50$, i.e., the proportion of DIF items remains the same. 

The intercept parameters $d_j$ are generated from the $\text{Uniform}(-2, 2)$ distribution and the slope parameters $a_j$ from the $\text{Uniform}(0.5, 1.5)$ distribution. In the two-group setting, we consider three cases for the class proportions, where $\nu = (0.1, 0.9)$, $\nu = (0.2, 0.8)$, and $\nu = (0.5, 0.5)$, respectively. The latent construct $\theta$ within each latent class $\xi$ follows
$$
\theta_i|\xi = 0 \sim \mathcal{N}(0, 1) \,\,\,\, \text{and} \,\,\,\, \theta_i|\xi = 1 \sim \mathcal{N}(0.5, 1.5),
$$ 
and for the three-group case, we additionally let $\theta_i|\xi = 2 \sim \mathcal{N}(1, 1.2)$. The DIF effect parameters are generated as $\delta_{jk} \sim \text{Uniform}(0.5, 1.5)$ for the non-zero elements and set to 0 for the remaining items. For the three-group case, the DIF effects for the second and third latent class are generated from a $\text{Uniform}(0.5, 1)$ and $\text{Uniform}(1, 1.5)$ distribution, respectively. In the two-group setting, the DIF items are positioned at the beginning of the scale (items 1-10 when $J=25$ and 1-20 when $J=50$). In the three-group setting, the DIF items are positioned at the end of the scale (items 16-25 when $J=25$ and 31-50 when $J=50$). The true model parameters are given in the supplementary material.
 
\subsection{Evaluation Criteria}

\paragraph{Detection of DIF items.} We check whether the DIF items can be accurately detected by the proposed method. In this analysis, we assume the number of latent classes $K$ is known. We calculate the average True Positive Rate (TPR) as 
\begin{equation}\label{eq:TPR}
 \widebar{\text{TPR}} = \frac{1}{B} \sum_{b=1}^B\frac{\sum_{j,k}\mathbbm{1}_{ \{\hat{\delta}^{(b)}_{jk} \neq 0, \delta_{jk}^{\!\scalebox{0.6}{*}} \neq 0 \}} }{\sum_{j,k} \mathbbm{1}_{\{\delta_{jk}^{\!\scalebox{0.6}{*}} \neq 0\}}},
\end{equation}
where  
$ \{ \hat{\delta}^{(b)}_{jk} \}_{J \times K}
$ 
is the estimated DIF effect matrix in the $b$-th replication and $\{\delta_{jk}^{\!\scalebox{0.6}{*}}\}_{J\times K}$ denotes the true DIF effect matrix. Similarly, we calculate the average true negative rate ($\widebar{\text{TNR}}$), which is the failure rate of identifying zero elements:
\begin{equation}\label{eq:TNR}
 \widebar{\text{FPR}} = \frac{1}{B} \sum_{b=1}^B \frac{\sum_{j,k}\mathbbm{1}_{ \{\hat{\delta}^{(b)}_{jk} \ne 0, \delta_{jk}^{\!\scalebox{0.6}{*}} = 0 \}} }{\sum_{j,k} \mathbbm{1}_{\{\delta_{jk}^{\!\scalebox{0.6}{*}} = 0\}}}.
\end{equation}
To better evaluate the performance of the proposed method in detecting DIF items, we compare the FPR and TPR with the results under an oracle setting where the group membership $\xi_i$ is known while anchor items are unknown. 
Under this oracle setting, we solve the following regularised estimation problem as in \citet{bauer2020simplifying}:
\begin{equation}\label{criterion2}
 \min_{\Xi}  -\log L^{ora}(\Xi) + \lambda \sum_{j=1}^J\sum_{k=1}^K |\delta_{jk}|,
\end{equation}
where 
$$L^{ora}(\Xi)  = \prod_{i=1}^N    \int \left(\prod_{j=1}^J \left(\exp((a_j\theta + d_j + \delta_{j\xi_i})Y_{ij})/(1+\exp(a_j\theta + d_j + \delta_{j\xi_i}))\right)\right) \phi(\theta \vert \mu_{\xi_i}, \sigma^2_{\xi_i}) d\theta$$
and $\Xi$ includes the parameters in $\Delta$ except for those in $\boldsymbol \nu$. The tuning parameter $\lambda$ is chosen by BIC. The FPR and TPR for detecting DIF items are also calculated under this setting and compared with those from the proposed method where $\xi_i$s are unknown. 

\paragraph{Classification of respondents.}
We then consider the classification of respondents based on the MAP estimate. Again, we assume the number of latent focal groups $K$ is known. We calculate the average classification error rate, given by the fraction of incorrect classifications to the overall number of classifications averaged over the number of replications:
$$
\text{error}_{MAP} = \frac{1}{N} \sum_{i=1}^N [\hat{\xi}_{\text{MAP},i} \neq \xi_i],
$$
where
$$
\hat{\xi}_{\text{MAP}, i} = \argmax_{k \in \{0, 1, ..., K\}} \hat{P}(\Hat{\xi} = k|\mathbf{y}_i)
$$
and $\hat{P}(\hat{\xi} = k | \mathbf{y}_i)$ is the posterior probability of category $k$ given in \eqref{eq:postprob}.
We notice that there is a label-switching problem under the setting with $K = 2$ when calculating the classification error. This problem is solved by a post-hoc label switching based on the estimated $\nu_1$ and $\nu_2$, using the ordering information that class 2 is larger than class 1, i.e.,  $\nu_2 >  \nu_1$. 
We also calculate the MAP classification error under the true model and compare it with the classification error of the proposed method.

\paragraph{Parameter estimation accuracy.} We further evaluate the accuracy of our final estimator $\hat{\Delta}^{(\hat \lambda)}$, assuming that $K$ is known. For each unknown parameter, the root mean square error (RMSE) and the absolute bias are calculated based on the 100 replications. 

\subsection{Results}
The classification performance in the simulation study is presented in Tables \ref{tab:accuracy} and \ref{tab:accuracy2}, which display the respondent and item classification accuracy for the two-group and three-group settings, respectively. First, it is observed that the classification error is predominantly influenced by the number of items, with larger item sizes resulting in better respondent classification performance. This observation is consistent with previous literature on DIF detection using IRT models \citep{chen2022detection, kuha2015nonequivalence}.

For respondent classification in the two-group setting, we observe in Table \ref{tab:accuracy} that the classification error is small for all simulation scenarios, and always better than a naïve classifier that assigns all respondents to the reference group. This is true even when the focal group only consists of 10\% of the respondents. As the proportion in the focal group (in Table \ref{tab:accuracy} denoted by $\pi$) increases, the proposed method's enhancement over the naïve classifier, which assigns all respondents to the baseline group, becomes more pronounced. We also note that the AUC values increase when the proportion of respondents in the focal group increases, but the increments are small. Table \ref{tab:accuracy} furthermore shows that the classification accuracy is only slightly worse when using the estimated parameters compared to when the true model parameters are used. 

For item classification in the two-group setting, the true positive rates are very high, and with no item being falsely flagged as a DIF item, across all scenarios. This suggests that the proposed framework is effective in identifying DIF items and minimizing false positives for various numbers of items and proportions in the focal group. We also note that the oracle estimator performance is only slightly better, i.e., knowing the group membership of the respondents only leads to marginal improvement in item classification accuracy.

In the three-group setting, Table \ref{tab:accuracy2} shows that the classification error rates are generally higher than those observed in the two-group case, which is expected given the increased complexity of the DIF detection problem when more than two groups are involved. However, note that the classification error is always clearly smaller than the naïve classifier that assigns all of the respondents to the reference group. This increase in classification performance is particularly clear in the simulation scenarios with $J=50$. We also observe that the AUC values for classes 2 and 3 are within reasonable ranges, suggesting that the proposed method is capable of adequately classifying respondents even in more challenging settings. The TPR and FPR values for item classification in the three-group setting are furthermore high for both class 2 and class 3 items and with no misclassified DIF-free item. This further supports the effectiveness of the proposed framework in identifying DIF items across different group configurations.

\begin{table}[]
\caption{Respondent and item classification accuracy under different simulation scenarios for the two-group case. The classification error and AUCs present respondent classification performance, where `AUC true' gives the results using the true parameter values. The TPRs and FPRs present the results for item classification, where `TPR oracle' and `FPR oracle' are the performance of the oracle estimator. }
\centering
\begin{tabular}{@{}cccccccc@{}}
\toprule
                     &                      &                      &                      &   $N=1,000$                    &               &                      &                      \\ \midrule
                     &                      & $J=25$                 &                      &                      &                      & $J=50$                 &                      \\ \cmidrule(lr){3-3} \cmidrule(lr){7-7}
        Evaluation measure              & $\pi=0.1$               & $\pi=0.2$               & $\pi=0.5$               &                     & $\pi=0.1$               & $\pi=0.2$               & $\pi=0.5$               \\ \cmidrule(lr){1-2} \cmidrule(lr){2-4} \cmidrule(l){6-8} 
Classification error & 0.078                & 0.141                & 0.256                &                      & 0.071                & 0.113                & 0.179                \\
Classification error true & 0.077   &  0.140      & 0.252   &  & 0.069   &     0.111   & 0.175   \\
AUC                  & 0.796                & 0.817                & 0.824                &                      & 0.896                & 0.903                & 0.904                \\
AUC true             & 0.813                & 0.827                & 0.830                &                      & 0.903                & 0.907                & 0.908                \\
\hdashline
TPR                  & 0.957                & 0.948                & 0.948                &                      & 0.997                & 0.993                & 0.996                \\
 FPR                  & 0                    & 0                    & 0                    &                      & 0                    & 0                    & 0                    \\
TPR oracle           &         1             &     1                 &     1                 &                      &    1                  &     1                 &    1                  \\
FPR oracle           &     0                 &       0               &     0                &                      &    0                  &    0                  &    0                  \\ \midrule
\multicolumn{1}{l}{} & \multicolumn{1}{l}{} & \multicolumn{1}{l}{} & \multicolumn{1}{l}{} & \multicolumn{1}{l}{} & \multicolumn{1}{l}{} & \multicolumn{1}{l}{} & \multicolumn{1}{l}{} \\ \midrule
                     &                      &                      &                      & $N=5,000$               &                      &                      &                      \\ \midrule
                     &                      & $J=25$                 &                      &                      &                      & $J=50$                 &                      \\ \cmidrule(lr){3-3} \cmidrule(lr){7-7}
            Evaluation measure         & $\pi=0.1$               & $\pi=0.2$               & $\pi=0.5$               &                      & $\pi=0.1$               & $\pi=0.2$               & $\pi=0.5$               \\ \cmidrule(lr){1-2} \cmidrule(lr){2-4} \cmidrule(l){6-8} 
Classification error & 0.087                &         0.161             &  0.259                    &                      &   0.074                   &          0.124            &      0.177                \\
Classification error true & 0.086   &  0.160      & 0.253   &  & 0.073   &     0.122   & 0.174   \\
AUC                  & 0.809                &      0.812                & 0.821                     &                      &   0.897                   &        0.901              &        0.904              \\
AUC true             & 0.819                &           0.819           &  0.827                    &                      &      0.902                &         0.904             &      0.907                \\
\hdashline
TPR                  & 0.953                &       0.947               &  0.950                    &                      &   0.995                   &         0.993             &       0.996               \\
FPR                  & 0                    &            0          &    0                  &                      &     0                 &     0                 &           0           \\
TPR oracle           &    1                  &     1                 &    1                  &                      &    1                  &     1                 &     1                 \\
FPR oracle           &  0                    &   0                  &     0                 &                      &     0                 &     0                 &      0                \\ \bottomrule
\end{tabular}
\label{tab:accuracy}
\end{table}

\begin{table}[]
\caption{Respondent and item classification accuracy under different simulation scenarios for the three-group case. The classification error and AUCs present respondent classification performance, where `AUC true' gives the results using the true parameter values. The TPRs and FPRs present the results for item classification, where `TPR oracle' and `FPR oracle' are the performance of the oracle estimator.}
\centering
\begin{tabular}{@{}ccccccc@{}}
\toprule
                     & \multicolumn{2}{c}{$N=1,000$} & \multicolumn{2}{c}{$N=5,000$} \\ 
\midrule
            Evaluation measure  & $J=25$ & $J=50$ & $J=25$ & $J=50$ \\
    \cmidrule(lr){1-2} \cmidrule(lr){2-3} \cmidrule(l){4-5} 
Classification error & 0.391  & 0.353  & 0.397  & 0.357  \\
Classification error true & 0.393 & 0.355 & 0.400 & 0.357 \\
AUC class 2          & 0.725  & 0.777  & 0.723  & 0.764  \\
AUC class 3          & 0.782  & 0.812  & 0.775  & 0.822  \\
AUC class 2 true     & 0.743  & 0.794  & 0.742  & 0.779  \\
AUC class 3 true     & 0.807  & 0.836  & 0.794  & 0.838  \\
\hdashline
TPR class 2          & 0.925  & 0.971  & 0.931  & 0.960  \\
TPR class 3          & 1      & 1      & 1      & 1      \\
FPR class 2          & 0      & 0      & 0      & 0      \\
FPR class 3          & 0      & 0      & 0      & 0      \\
TPR class 2 oracle   & 1      & 0.95   & 1      & 1      \\
TPR class 3 oracle   & 1      & 1      & 1      & 1      \\
FPR class 2 oracle   & 0      & 0      & 0      & 0      \\
FPR class 3 oracle   & 0      & 0      & 0      & 0      \\ 
\bottomrule
\end{tabular}
\label{tab:accuracy2}
\end{table}

\begin{figure}
    \centering
    \includegraphics[scale=0.9]{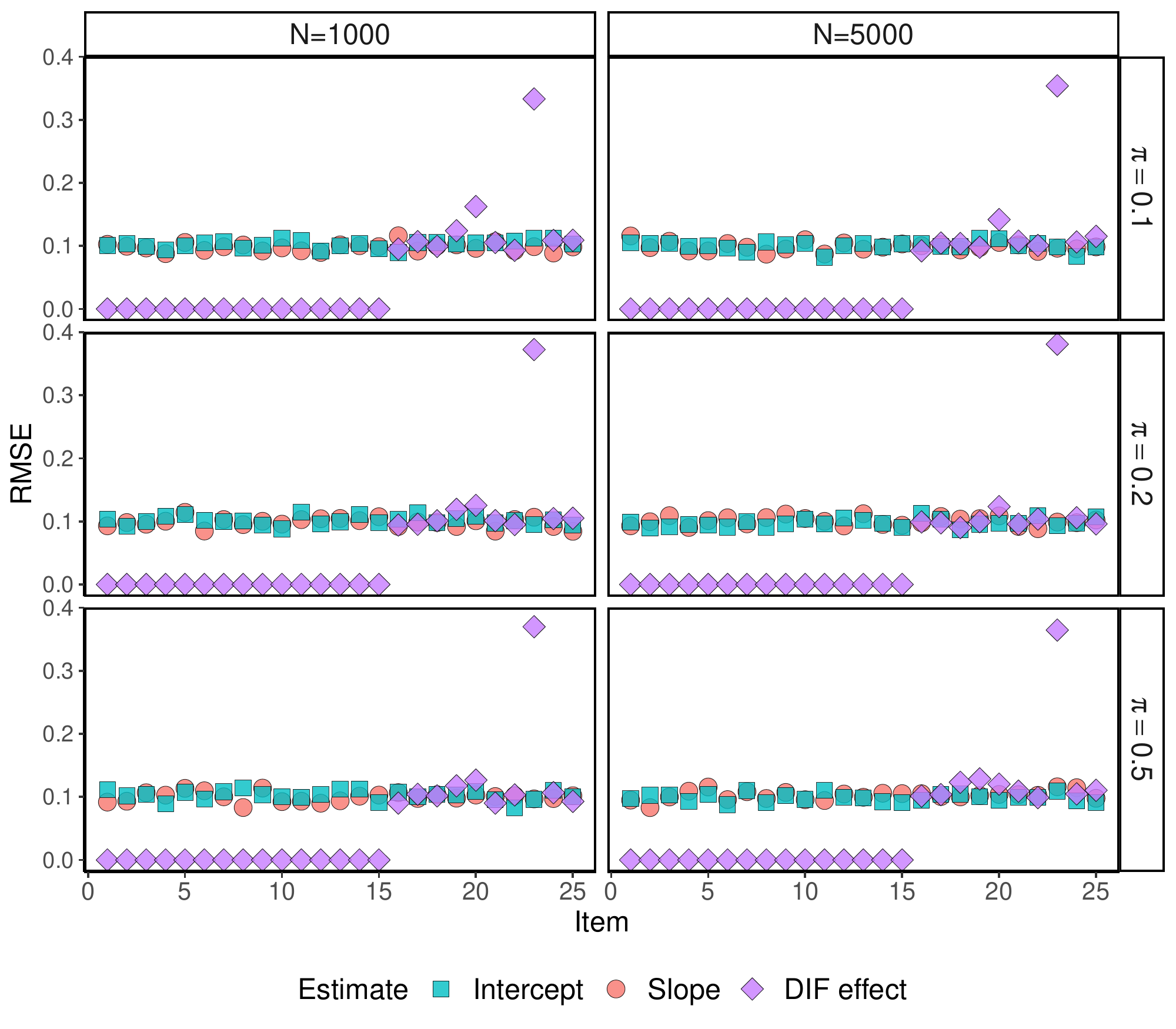}
    \caption{RMSE for $J=25$ under the 2-group setting.}
    \label{fig:RMSE_J25}
\end{figure}

\begin{figure}
    \centering
    \includegraphics[scale=0.9]{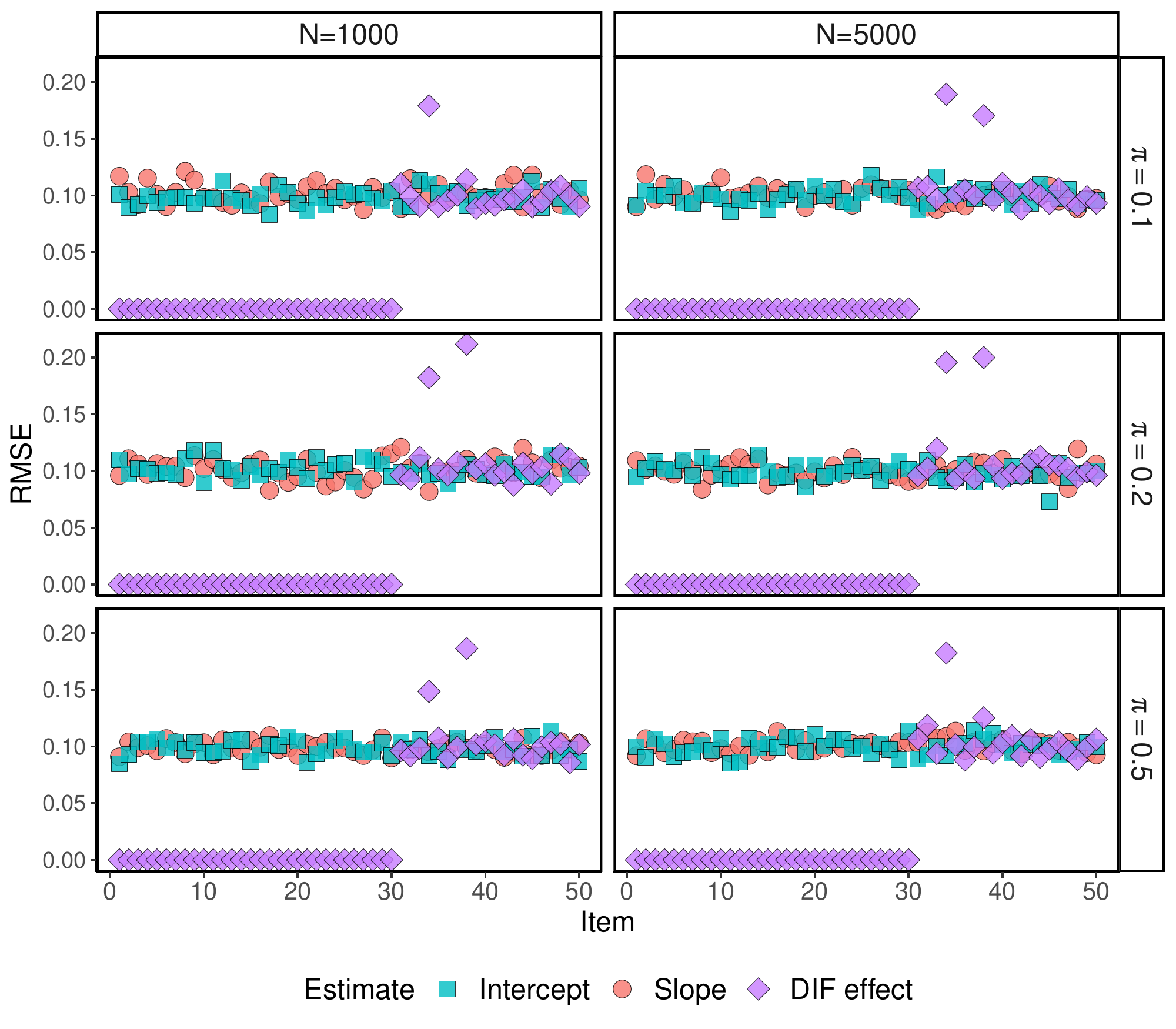}
    \caption{RMSE for $J=50$ under the 2-group setting.}
    \label{fig:RMSE_J50}
\end{figure}

Figures \ref{fig:RMSE_J25} and \ref{fig:RMSE_J50} show the RMSEs (Root Mean Squared Errors) for all the item parameter estimates for the two-group setting ($K=1$). The RMSEs are small for all estimates across the items, with the exemption of one or two items that show larger RMSEs for the estimated DIF parameter. The RMSEs for the estimated DIF parameters for the DIF-free items are zero as the proposed estimation procedure successfully classifies the DIF items. We also see that the number of items and proportion of respondents in the focal group have essentially no influence on the RMSE, for the configurations considered in this simulation study.

\begin{figure}
    \centering
    \includegraphics[scale=0.9]{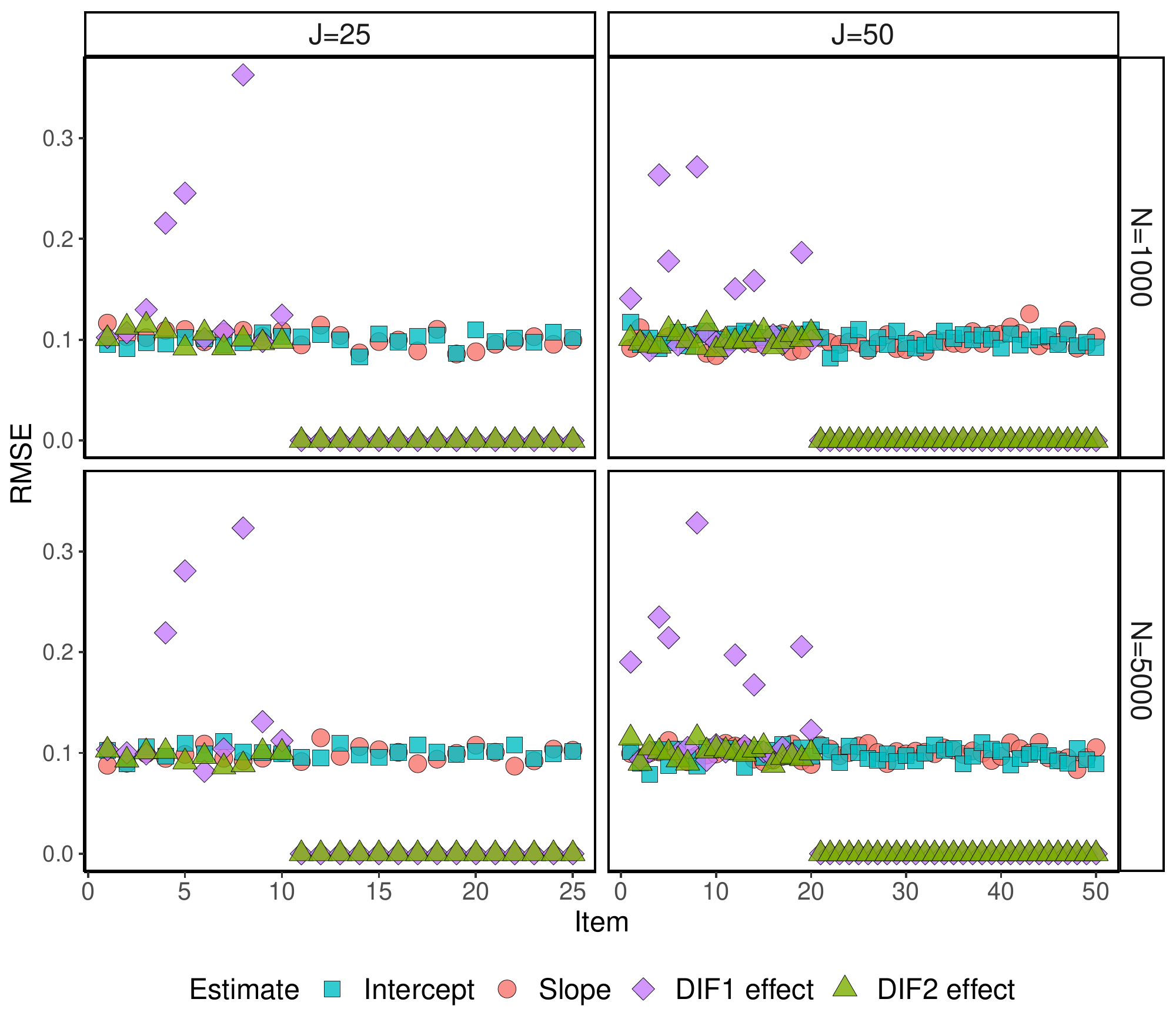}
    \caption{The RMSEs for under the 3-group setting.}
    \label{fig:RMSE_3group}
\end{figure}

In Figure \ref{fig:RMSE_3group}, we can observe the RMSEs specifically for the three-group scenario. It is important to note that in this case, there exist two DIF effects, one for each focal group. For the first focal group, the true DIF effects are drawn from values in the range [0.5 -- 1] and for the second focal group, they are drawn from the range [1 -- 1.5]. The increased difficulty of estimating smaller DIF effects is reflected by larger RMSEs for the first focal group. 

In Table \ref{tab:paramJ25} and Table \ref{tab:paramJ50}, the absolute bias and RMSE under the 2-group setting, averaged over the number of items of the same type, are displayed for both sample sizes and every focal group proportion $\pi$ considered. The absolute bias and RMSE are small for all parameters, and differences are very small between different values of $\pi$ and different values of $N$. The most notable difference is seen for the estimate of the focal group proportion $\pi$, where the bias and RMSE clearly decrease when the sample size increases. We also see that the DIF effect parameter is estimated with only a small bias and RMSE.
In Table \ref{tab:RMSE_3group}, the absolute bias and RMSE are shown for the 3-group setting. As in the two-group setting, the bias and RMSE values are small under all settings.
\begin{table}[]
\centering
\caption{Average absolute bias and RMSE over all items by estimated parameter type, when \(J=25\), \(N=1000\), and \(5000\), under the 2-group setting.}
\begin{tabular}{@{}cccccc@{}}
\toprule
         &           & \multicolumn{2}{c}{\(N=1,000\)} & \multicolumn{2}{c}{\(N=5,000\)} \\ \midrule
Scenario & Parameter & Abs. bias        & RMSE       & Abs. bias        & RMSE       \\ \midrule
\multirow{6}{*}{\(\pi=0.1\)} & \(d\)         & 0.078          & 0.098      & 0.079            & 0.099 \\
         & \(a\)         & 0.082          & 0.103      & 0.081            & 0.100           \\
         & \(\delta\)   & 0.040          & 0.053      & 0.041            & 0.053           \\
         & \(\mu\)       & 0.087          & 0.114      & 0.077            & 0.098           \\
         & \(\sigma\)    & 0.074          & 0.093      & 0.088            & 0.108           \\
         & \(\pi\)       & 0.035          & 0.040      & 0.017            & 0.022           \\
\hdashline
\multirow{6}{*}{\(\pi=0.2\)} & \(d\)         & 0.078          & 0.098      & 0.081            & 0.101           \\
         & \(a\)         & 0.081          & 0.101      & 0.078            & 0.098           \\
         & \(\delta\)   & 0.041          & 0.053      & 0.041            & 0.052           \\
         & \(\mu\)       & 0.081          & 0.105      & 0.082            & 0.107           \\
         & \(\sigma\)    & 0.082          & 0.101      & 0.086            & 0.106           \\
         & \(\pi\)       & 0.050          & 0.056      & 0.022            & 0.027           \\
\hdashline
\multirow{6}{*}{\(\pi=0.5\)} & \(d\)         & 0.079          & 0.100      & 0.081            & 0.102           \\
         & \(a\)         & 0.082          & 0.103      & 0.079            & 0.099           \\
         & \(\delta\)   & 0.040          & 0.052      & 0.042            & 0.055           \\
         & \(\mu\)       & 0.084          & 0.104      & 0.076           & 0.093           \\
         & \(\sigma\)    & 0.075          & 0.094      & 0.080            & 0.101           \\ 
         & \(\pi\)        & 0.040          & 0.049      & 0.030            & 0.037           \\
\bottomrule
\end{tabular}
\label{tab:paramJ25}
\end{table}

\begin{table}[]
\centering
\caption{Average absolute bias and RMSE over all items by estimated parameter type, when \(J=50\), \(N=1000\), and \(5000\) under the 2-group setting.}
\begin{tabular}{@{}cccccc@{}}
\toprule
         &           & \multicolumn{2}{c}{\(N=1,000\)} & \multicolumn{2}{c}{\(N=5,000\)} \\ \midrule
Scenario & Parameter & Abs. bias        & RMSE       & Abs. bias        & RMSE       \\ \midrule
\multirow{6}{*}{\(\pi=0.1\)} & \(d\)         & 0.082          & 0.102      & 0.080            & 0.100           \\
         & \(a\)         & 0.078          & 0.098      & 0.079            & 0.100           \\
         & \(\delta\)   & 0.032          & 0.041      & 0.033            & 0.043           \\
         & \(\mu\)       & 0.081          & 0.102      & 0.079            & 0.096           \\
         & \(\sigma\)    & 0.078          & 0.099      & 0.077            & 0.096           \\
         & \(\pi\)       & 0.026          & 0.030      & 0.013            & 0.016           \\
\hdashline
\multirow{6}{*}{\(\pi=0.2\)} & \(d\)         & 0.081          & 0.102      & 0.080            & 0.100            \\
         & \(a\)         & 0.081          & 0.102      & 0.079            & 0.099           \\
         & \(\delta\)   & 0.034          & 0.044      & 0.034            & 0.044           \\
         & \(\mu\)       & 0.081          & 0.100      & 0.077            & 0.101           \\
         & \(\sigma\)    & 0.082          & 0.097      & 0.084            & 0.106           \\
         & \(\pi\)       & 0.034          & 0.038      & 0.014            & 0.018           \\
\hdashline
\multirow{6}{*}{\(\pi=0.5\)} & \(d\)         & 0.079          & 0.099      & 0.081            & 0.101           \\
         & \(a\)         & 0.079          & 0.098      & 0.080            & 0.100           \\
         & \(\delta\)   & 0.032          & 0.042      & 0.032            & 0.042           \\
         & \(\mu\)       & 0.091          & 0.112      & 0.078            & 0.097           \\
         & \(\sigma\)    & 0.073          & 0.093      & 0.081            & 0.106           \\ 
         & \(\pi\)       & 0.030          & 0.038      & 0.021            & 0.025           \\
\bottomrule
\end{tabular}
\label{tab:paramJ50}
\end{table}

\begin{table}[]
\centering
\caption{Average absolute bias and RMSE over all items by estimated parameter type, \(N=1000\) and \(5000\), under the 3-group setting.}
\begin{tabular}{@{}cccccc@{}}
\toprule
         &           & \multicolumn{2}{c}{\(N=1,000\)} & \multicolumn{2}{c}{\(N=5,000\)} \\ \midrule
Scenario & Parameter & Abs. bias        & RMSE       & Abs. bias        & RMSE       \\ \midrule
\multirow{10}{*}{\(J=25\)} & \(d\)         & 0.081      & 0.101      & 0.078 & 0.098                    \\
         & \(a\)          & 0.080          & 0.100      & 0.080            & 0.101                      \\
         & \(\delta_2\)   & 0.046          & 0.064      & 0.045          & 0.062                  \\
        & \(\delta_3\)    & 0.033          & 0.041      & 0.031            & 0.038                  \\
         & \(\mu_2\)      & 0.084          & 0.103      & 0.079            & 0.095                   \\
         & \(\mu_3\)      & 0.074          & 0.094     &  0.073           & 0.094                     \\
         & \(\sigma_2\)   & 0.085          & 0.107      & 0.083            & 0.103                    \\
         & \(\sigma_3\)   & 0.081          & 0.099      & 0.084            & 0.103                      \\
         & \(\pi_2\)      & 0.060          & 0.075      & 0.045            & 0.057                  \\
         & \(\pi_3\)      & 0.045          & 0.057      & 0.044            & 0.054                 \\
\hdashline
\multirow{10}{*}{\(J=50\)} & \(d\)         & 0.080      & 0.099      & 0.079 & 0.100 \\
         & \(a\)          & 0.08          & 0.101      & 0.078            & 0.098           \\
         & \(\delta_2\)   & 0.038          & 0.052      & 0.039            & 0.057           \\
        & \(\delta_3\)    & 0.033          & 0.040      & 0.032            & 0.040           \\
         & \(\mu_2\)      & 0.083          & 0.100      & 0.083            & 0.104           \\
         & \(\mu_3\)      & 0.072          & 0.090      & 0.080            & 0.097           \\
         & \(\sigma_2\)   & 0.076          & 0.092      & 0.085            & 0.107           \\
         & \(\sigma_3\)   & 0.081          & 0.103      & 0.072            & 0.096           \\
         & \(\pi_2\)      & 0.064          & 0.079      & 0.055            & 0.067           \\
         & \(\pi_3\)      & 0.044          & 0.056      & 0.035            & 0.046           \\
\bottomrule
\end{tabular}
\label{tab:RMSE_3group}
\end{table}

In summary, the simulation results presented in Tables \ref{tab:accuracy}--\ref{tab:RMSE_3group}, and Figures \ref{fig:RMSE_J25}--\ref{fig:RMSE_3group} demonstrate the potential of the proposed framework for DIF analysis with unknown anchor items and comparison groups. The framework performs well in terms of respondent and item classification accuracy across a range of scenarios, and with good parameter recovery, suggesting its applicability in various real-world settings.

\section{Real Data Analysis}\label{EmpiricalAnalysis}

To illustrate the proposed method, we analyse a mathematics test from a midwestern university in the United States. This data set has been analysed in both \citet{bolt2002item} and \citet{de2011explanatory}. The data contains 3000 examinees answering 26 binary-scored items. The original dataset contains two test forms, with 8 items in common.\footnote{The original dataset contains 36 items, with 12 common items. We follow the procedure in \citet{bolt2002item} and only analyse items 1-18 and 29-36. In this analysis, items 19-26 therefore actually occupied item position 29-36.}. Six of the common items are of particular interest as they are positioned at the end of the test. \citet{bolt2002item} hypothesised the existence of two latent classes: one speeded class that answered end-of-test items with insufficient time, and another non-speeded class. The identification of speeded items was conducted using a two-form design. Specifically, they examined common items across two test administrations, where the common items were placed at the end of the test in one form and earlier in the other form. By estimating the item difficulty for the end-of-test common items and comparing it to the difficulty estimates from the other form, they were able to quantify the DIF effect. Our goal is to detect the DIF items, i.e., the speeded items, and classify respondents into latent classes. Thanks to our procedure, we can analyse only one form, without using information from the second form. A similar analysis using simulated data was also conducted in \cite{robitzsch2022regularized} but using a Rasch mixture model.

We start by fitting the proposed model to the data for different values of $K$, which determines the number of latent classes. The BIC for a model with $K=0$, i.e. no latent classes other than the reference group, equals 117,552.2, the BIC for $K=1$ equals 92,300.1, and for $K=2$, the BIC equals 92,522.8. We therefore proceed with a model using $K=1$, i.e., two latent classes. This aligns with the two-group model considered in \citet{bolt2002item}. It took the proposed EM algorithm 37.04, 144.760, and 277.390 seconds to converge\footnote{In our implementation of the proposed EM algorithm, we stop the iterations when the increase in the log-likelihood is smaller than $10^{-4}$.} for the 1-. 2-, and 3-group solutions, respectively\footnote{CPU configuration: 11th Gen Intel(R) Core(TM) i7-1165G7 2.80GHz 3200MHz}. The proposed model classifies 25.8\% into the second latent. If we interpret the two classes as a speeded and non-speeded class, this means that about 26\% of the respondents belong to the speeded class. The estimated mean ability in the speeded class equals -0.351 with the estimated standard deviation equal to 1.075. Since the reference group (the non-speeded class) has a prespecified ability mean and standard deviation equal to 0 and 1, respectively, our results therefore indicate that the speeded class has a lower ability on average compared to the non-speeded class. These findings align closely with the results presented in \citet{bolt2002item}.

In Table \ref{table:1} we give the estimated item parameters from the educational test data. The estimated item discrimination and easiness parameters, $\hat{a}$ and $\hat{d}$ respectively, are provided together with the estimated DIF effect $\hat{\delta}$. Common items are denoted by asterisks. For the majority of the items, the DIF effects are estimated to be zero, indicating that these items do not exhibit any significant measurement bias between different groups. However, items 20-26 exhibit non-zero DIF effects, suggesting that these items might be functioning differently for the two latent classes. Among these, items 20, 21, 22, 23, and 24 are also common items, which may require further investigation to ensure fair assessment across test administrations. Since the DIF effect for these end-of-test items is all negative, it suggests that these items become more difficult for the second latent class. This class could therefore consist of respondents that ran out of time and had insufficient time to answer these items. This is known as a speededness effect. As a result, the item difficulty is inflated, which could lead to biased subsequent analyses. The presence of non-zero DIF effects for some end-of-test items highlights the need to scrutinize these items more closely and potentially revise the test administration to minimize the impact of speededness. For instance, increasing the allocated time for the test or redistributing the items more evenly throughout the test could help alleviate the speededness effect and create a more unbiased assessment.

\begin{table}[t]
\makebox[\textwidth][c]{ 
\begin{tabular}{cccc}
\hline
Item & $\hat{a}$ & $\hat{d}$ & $\hat{\delta}$ \\ \hline
1 & 1.298 & 2.987 & 0 \\
2 & 1.287 & 1.584 & 0 \\
3 & 0.552 & 2.968 & 0 \\
4 & 0.878 & 0.707 & 0 \\
5 & 1.336 & 2.247 & 0 \\
6 & 0.973 & 0.822 & 0 \\
7 & 0.669 & 0.296 & 0 \\
8 & 0.687 & -0.882 & 0 \\
9 & 1.268 & 0.908 & 0 \\
10 & 0.874 & 2.101 & 0 \\
11* & 0.984 & -0.575 & 0 \\
12 & 0.912 & -0.804 & 0 \\
13 & 1.040 & 0.917 & 0 \\ \hline
\end{tabular}
\hspace{0.5cm} 
\begin{tabular}{cccc}
\hline
Item & $\hat{a}$ & $\hat{d}$ & $\hat{\delta}$ \\ \hline
14 & 0.746 & 0.941 & 0 \\
15 & 0.471 & -1.646 & 0 \\
16* & 1.507 & -0.526 & 0 \\
17 & 1.271 & 1.843 & 0 \\
18 & 1.249 & -2.588 & 0 \\
19 & 1.071 & 0.944 & 0 \\
20* & 1.017 & 0.455 & -1.411 \\
21* & 1.425 & 1.324 & -1.808 \\
22* & 0.664 & -0.608 & -1.104 \\
23* & 0.767 & -0.063 & -0.624 \\
24* & 0.929 & -0.639 & -0.885 \\
25 & 1.247 & -2.020 & -1.019 \\
26* & 1.340 & -0.079 & -0.851 \\ \hline
\end{tabular}
}
\caption{Estimated item easiness and DIF effects for the detected DIF items. The asterisks denote the common items.}
\label{table:1}
\end{table}

\section{Concluding Remarks}\label{ConcludingRemarks}

In this paper, we presented a comprehensive framework for DIF analysis that overcomes several limitations of existing methods. Our approach can deal with the situation in which both anchor items and comparison groups are unknown, a setting commonly encountered in real-world applications. 
The use of latent classes in our approach allows us to model heterogeneity among the observations. In this sense, our approach relates to an exploratory dimensionality analysis where there is, in addition to the primary latent dimension, a second dimension that is treated as unknown. In our empirical analysis, this additional dimension is labeled as a speededness effect. In addition to modeling the additional latent dimension(s), the proposed regularised estimator enables us to identify DIF items and quantify their effect on the intercept parameter of the model. We also propose an efficient EM algorithm for the estimation of the model parameters\footnote{The R code for the proposed method is available from https://github.com/gabrieltwallin/LatentDIF/}. One merit of our framework is its flexibility. While focusing on the 2-PL model as the baseline model, our approach can be easily extended to accommodate other widely used IRT models, such as the Rasch model and the proportional odds model. We can also allow the baseline model to be a multidimensional IRT model, as shown in Section \ref{sec:Extensions}. Our framework is furthermore able to accommodate more than two comparison groups, allowing DIF effects to vary between the groups. Lastly, the proposed method can be extended as shown in Section 2.6 to detect non-uniform DIF. This flexibility makes our framework applicable to a wide range of contexts. 

Although our approach shows promising results, there are still several limitations to be addressed in future research. For example, we do not provide confidence intervals for the DIF effect parameters which would be useful for practitioners and researchers in interpreting the magnitude and significance of the DIF effects. In \citet{chen2021dif} for example, where the comparison groups are known but the anchor items are unknown, the distribution of $\hat{\delta}_{jk} - \delta_{jk}$ is approximated by Monte Carlo simulation to yield valid statistical inference. This procedure does in essence apply to our case as well. In addition, we have not linked the latent classes to covariates, as in for example \citet{vermunt2010latent} and \citet{vermunt2021perform}. By doing so, researchers can gain insights into the underlying characteristics of the different classes and better understand the factors that may contribute to DIF. This would enhance the interpretability of the results and help identify potential sources of DIF that could be addressed in the development of assessment instruments. To address this limitation, future research could explore the integration of covariates within a structural equation modeling (SEM) framework. This would enable the simultaneous modelling of both the measurement model (i.e., the IRT model) and the structural model (i.e., relationships between latent variables and covariates). Incorporating covariates in this manner would not only improve the interpretability of the results but could also provide a more comprehensive understanding of the relationships between the items, latent traits, and potential sources of DIF. Our framework could also be extended to accommodate non-uniform DIF, i.e., DIF in the slope parameter, such as in \citet{wang2021using} that considers a multidimensional IRT model with known comparison groups and unknown anchor items.

In this study, we focus on the Lasso penalty for its simplicity, computational efficiency, and well-documented ability to perform both variable selection and regularisation. The Lasso's convex optimisation problem is easier to solve computationally than some non-convex penalties like the SCAD \citep[][]{fan2001variable} and the Minimax Concave Penalty \citep[MCP;][]{zhang2010nearly}. We acknowledge that the Lasso penalty can introduce some bias into parameter estimates. However, in our proposed method we use the Lasso for model selection. As we thereafter refit the selected model there will be no bias, asymptotically, supposing that the model selection based on the Lasso is consistent \citep{zhao2021defense}. Alternative penalties, including the adaptive Lasso \citep{zou2006adaptive}, SCAD, and MCP, have their own merit. However, they also come with some challenges, especially in terms of computational complexity and algorithm stability. We, therefore, argue that the Lasso penalty is a suitable choice for the proposed model and its identifying assumptions. We believe it would be interesting in the future to compare the performance of estimators with different penalty functions under the current latent DIF setting. 

Our proposed framework provides a powerful tool for DIF analysis with unknown anchor items and comparison groups. The framework has the potential to inform the development of fair and unbiased assessments. Future research can build upon our approach by addressing the limitations and exploring other applications. In terms of the potential impact of our work, the framework could be particularly beneficial in specific contexts, such as educational assessment, where identifying and addressing DIF is critical to ensure that tests fairly measure students' abilities across heterogeneous populations, thereby promoting equal access to educational opportunities. It could also be considered in employment selection, where unbiased assessment instruments are crucial to creating a diverse and inclusive workforce that complies with legal requirements related to fairness in employment practices \citep{ployhart2008diversity, hough2001determinants}. Another application is psychological evaluations, where accurate identification of DIF can help improve diagnostic tools and treatment recommendations, leading to better outcomes for individuals from diverse backgrounds \citep{teresi2021differential}. By addressing the limitations and further refining our approach, this framework has the potential to contribute to the development of more fair assessment practices in these and other domains, ultimately benefiting a wide range of stakeholders.

\clearpage

\appendix

\section{Gradients for the Proximal Gradient Descent}

In the M-step of the proposed EM-algorithm, we implement a proximal gradient descent. This algorithm requires the gradients of the objective function, which for the proposed model can be expressed as
\begin{equation}\label{eq:gradients}
  \frac{\partial l(\Delta)}{\partial \eta_j} = \sum_{i=1}^N \int \frac{\partial \varphi_{ij}}{\partial \eta_j} \bigg[ y_{ij} - P(Y_{ij} = 1 | \theta_i, \xi_i) \bigg] \phi(\theta_i; \mu_k, \sigma^2_k) d \theta_i, 
\end{equation}
where $\eta_j$ is a generic notation for the item parameters $(a_j, d_j, \delta_{jk})$, and $\varphi_{ij} = d_j + a_j\theta_i + \delta_{j\xi_i}$, 

To give an example, we give the explicit expressions for the two-group case. We start by parametrising the latent construct of the focal group as $\theta_2 = \mu_2 + \sigma_2 \theta_1$, where $\mu$ and $\sigma$ is the mean and standard deviation of the latent construct in the focal group, and $\theta_1 \sim \mathcal{N}(0,1)$ is the latent construct in the reference group. We have that $\varphi_{ij}^{(1)} = d_j + a_j \theta_{i,1}$ for the reference group and $\varphi_{ij}^{(2)} = d_j + \delta_j + a_j \theta_{i,2}$ for the focal group. The partial derivatives in \eqref{eq:gradients} are given by
\begin{equation}
\begin{aligned}
    \frac{\partial \varphi_{ij}^{(1)}}{\partial d_j} &= 1 & \frac{\partial \varphi_{ij}^{(2)}}{\partial d_j} &= 1 \\
    \frac{\partial \varphi_{ij}^{(1)}}{\partial a_j} &= \theta_1 & \frac{\partial \varphi_{ij}^{(2)}}{\partial a_j} &= \mu_2 + \sigma_2 \theta_{i,1} \\
    \frac{\partial \varphi_{ij}^{(1)}}{\partial \delta_{j2}} &= 0 & \frac{\partial \varphi_{ij}^{(2)}}{\partial \delta_{j2}} &= 1 \\
    \frac{\partial \varphi_{ij}^{(1)}}{\partial \mu_2} &=  0 & \frac{\partial \varphi_{ij}^{(2)}}{\partial \mu_2} &=  1\\
    \frac{\partial \varphi_{ij}^{(1)}}{\partial \sigma_2} &= 0 & \frac{\partial \varphi_{ij}^{(2)}}{\partial \sigma_2} &= \theta_1 
\end{aligned}
\end{equation}

\clearpage

\section{The Line Search Procedure}

The implemented line search algorithm attempts to find an appropriate step size for the proximal gradient descent implemented in the M-step of the proposed EM algorithm. It does so by iteratively adjusting the step size until the change in the objective function (i.e., the penalised log-likelihood) is within a specified tolerance. The algorithm starts with an initial step size and iteratively reduces it by a factor (in this case, dividing by 2) until the new objective function value satisfies the tolerance condition. If the maximum number of iterations is reached without finding a satisfactory step size, the algorithm returns the current step size. The implemented line search algorithm is given in the following Line Search Algorithm.

\begin{algorithm}
\renewcommand{\thealgorithm}{}
\caption{Line Search Algorithm}\label{alg:LineSearch}
\textbf{Input:} $Y$, $a^{(0)}$, $d^{(0)}$, $\delta^{(0)}$, $\pi^{(0)}$, $\mu^{(0)}$, $\sigma^{(0)}$, $\nabla_a$, $\nabla_d$, $\nabla_\delta$, $\nabla_\mu$, $\nabla_\sigma$, $\lambda$, $\gamma^{(0)} = 1$.

\medskip
Initialize the objective function $$O^{(0)} = -l(\Delta^{(0)}) + \lambda \sum_{k=0}^K \sum_{j=1}^J |\delta^{(0)}_{jk}|$$ \\
Set the tolerance $tol$ to $10^{-6}$ \\
Set maximum number of iterations $T$ to 100\\

\medskip
\begin{enumerate}
\item[] \textbf{For} $t = 1, \ldots, T$ \textbf{do}

\begin{itemize}
    \item[] Update the parameters using the current step size:
    \begin{itemize}
        \item[] $a^{(t)} \gets a^{(t-1)} - \gamma^{(t-1)} \nabla_a$
        \item[] $d^{(t)} \gets d^{(t-1)} - \gamma^{(t-1)} \nabla_d$
        \item[] $\delta^{(t)} \gets \text{Prox}(\delta^{(t-1)}, \nabla_\delta, \lambda, \gamma^{(t-1)})$
        \item[] $\sigma^{(t)} \gets \sigma^{(t-1)} - \gamma^{(t-1)} \nabla_\sigma$
        \item[] $\mu^{(t)} \gets \mu^{(t-1)} - \gamma^{(t-1)} \nabla_\mu$
    \end{itemize}
    
    \item[] Calculate new objective function value $$O^{(t)} \gets -l(\Delta^{(t)}) + \lambda \sum_{k=0}^K \sum_{j=1}^J |\delta^{(t)}_{jk}|$$
    
    \item[] \textbf{If} $|O^{(t)} - O^{(t-1)}| < tol$ \textbf{then}
    \begin{itemize}
        \item[] \textbf{return} $\gamma^{(t)}$
    \end{itemize}
    \textbf{else}
    \begin{itemize}
        \item[] Adjust the step size and update the value of the objective function:
        \begin{itemize}
            \item[] $\gamma^{(t)} \gets \gamma^{(t-1)} / 2$
        \end{itemize}
    \end{itemize}
    \textbf{end if}
\end{itemize}
\item[] \textbf{end for}
\end{enumerate}
\medskip
\begin{itemize}
    \item[] $\gamma^{*} \gets \gamma^{(t)}$
\end{itemize}
\textbf{return} Optimal step size $\gamma^{*}$
\label{algo:line_search}
\end{algorithm}
\renewcommand{\thealgorithm}{\arabic{algorithm}}

\newpage
\section{The Soft-Thresholding Procedure}

The soft threshold function and the proximal gradient function are used to identify the DIF-free items, i.e., the anchor items, from the data. The soft threshold function takes a vector $x$ and a scalar $\lambda$ as inputs and applies element-wise thresholding to $x$. It sets elements with absolute values less than or equal to $\lambda$ to zero, subtracts $\lambda$ from elements greater than $\lambda$, and adds $\lambda$ to elements less than $-\lambda$. The proximal gradient function takes the (estimated) DIF effect $\delta$, its gradient $\nabla_x$, a regularisation parameter $\lambda$, and a step size $\gamma$ as inputs. It calls the soft threshold function with the updated vector $\delta - \gamma \nabla_\delta$ and the product $\lambda \gamma$. The output of the proximal gradient function is the updated DIF effect parameter estimate after applying the soft threshold function. These algorithms are summarised below.

\begin{algorithm}
\renewcommand{\thealgorithm}{}
\caption{Soft Threshold Function}\label{alg:SoftThreshold}
\textbf{Input:} $x$, $\lambda$

\medskip
\begin{enumerate}
\item[] Create a temporary variable $temp \gets x$
\item[] Set elements of $temp$ to 0 if their absolute value is less than or equal to $\lambda$: 
\begin{itemize}
\item[] $temp[|x| \leq \lambda] \gets 0$
\end{itemize}
\item[] Update elements of $temp$ where $x > \lambda$: 
\begin{itemize}
\item[] $temp[x > \lambda] \gets temp[x > \lambda] - \lambda$
\end{itemize}
\item[] Update elements of $temp$ where $x < -\lambda$: 
\begin{itemize}
\item[] $temp[x < -\lambda] \gets temp[x < -\lambda] + \lambda$
\end{itemize}
\end{enumerate}

\medskip
\textbf{Output:} $temp$
\end{algorithm}

\begin{algorithm}
\renewcommand{\thealgorithm}{}
\caption{Proximal Gradient Function}\label{alg:ProximalGradient}
\textbf{Input:} $\delta$, $\nabla_\delta$, $\lambda$, $\gamma$

\medskip
\begin{enumerate}
\item[] Call the soft threshold function with input $(\delta - \gamma \nabla_{\delta}, \lambda \gamma)$:
\begin{itemize}
\item[] $\text{Prox} \gets soft\_thre(\delta - \gamma \nabla_\delta, \lambda \gamma)$
\end{itemize}
\end{enumerate}

\medskip
\textbf{Output:} $\text{Prox}$
\end{algorithm}
\renewcommand{\thealgorithm}{\arabic{algorithm}}

\newpage
\section{The Closed-Form Solution of the Latent Class Proportion}

Using the Lagrange multiplier method, we obtain 
$$
\nu_k^{(t+1)} = \frac{\sum_{i=1}^n \gamma_{ik}^{(t)}}{n},
$$
where $\gamma_{ik}^{(t)} = P(\xi_i = k | \mathbf{y}_i)$

We can see this from the following argument. Given that
\begin{align*}
D_t(\Delta_1) = -\sum_{i=1}^N \mathbb{E}\left[ \log\left(\nu_{\xi_i} \big\vert \mathbf Y_i, \Delta^{(t)}\right)\right].
\end{align*}
The expectation can be expressed as
\begin{align*}
\mathbb{E}\left[\log(\nu_{\xi_i}) \big\vert \mathbf Y_i, \Delta^{(t)}\right] = \sum_{k=0}^K P(\xi_i = k \big\vert \mathbf Y_i, \Delta^{(t)}) \log(\nu_k).
\end{align*}
Thus, we have
\begin{align*}
D_t(\Delta_1) = -\sum_{i=1}^N \sum_{k=0}^K \gamma_{ik}^{(t)} \log(\nu_k),
\end{align*}
where $\gamma_{ik}^{(t)} = P(\xi_i = k \big\vert \mathbf Y_i, \Delta^{(t)})$.

Construct the Lagrangian function with the constraints:
\begin{align*}
\mathcal{L}(\Delta_1, \boldsymbol{\alpha}, \beta) = D_t(\Delta_1) + \sum_{k=0}^K \alpha_k (\nu_k - \epsilon_k) + \beta\left(\sum_{k=0}^K \nu_k - 1\right),
\end{align*}
where $\alpha_k \geq 0$ for $k = 0, 1, \dots, K$.

Now, take the partial derivative of the Lagrangian function with respect to $\nu_k$:
\begin{align*}
\frac{\partial \mathcal{L}}{\partial \nu_k} = -\sum_{i=1}^n \gamma_{ik}^{(t)}\frac{1}{\nu_k} + \alpha_k + \beta, \quad k = 0, 1, \dots, K.
\end{align*}

Setting the partial derivatives to zero, we get
\begin{align*}
-\sum_{i=1}^n \gamma_{ik}^{(t)}\frac{1}{\nu_k} + \alpha_k + \beta = 0, \quad k = 0, 1, \dots, K.
\end{align*}

From the above equation, we have:
\begin{align*}
\nu_k^{(t+1)} = \frac{\sum_{i=1}^n \gamma_{ik}^{(t)}}{-\alpha_k - \beta}, \quad k = 0, 1, \dots, K.
\end{align*}

To find the values of $\alpha_k$ and $\beta$, we need to apply the constraint $\sum_{k=0}^K \nu_k = 1$. Plugging in the expression for $\nu_k^{(t+1)}$, we have
\begin{align*}
\sum_{k=0}^K \frac{\sum_{i=1}^n \gamma_{ik}^{(t)}}{-\alpha_k - \beta} = 1.
\end{align*}

Since the constraint only involves the sum of the $\nu_k^{(t+1)}$, we can eliminate the Lagrange multipliers $\alpha_k$ by normalizing the solution:

\begin{align*}
\nu_k^{(t+1)} = \frac{\sum_{i=1}^n \gamma_{ik}^{(t)}}{\sum_{k=0}^K \sum_{i=1}^n \gamma_{ik}^{(t)}}, \quad k = 0, 1, \dots, K.
\end{align*}

However, since $\sum_{k=0}^K \gamma_{ik}^{(t)} = 1$ for all $i = 1, 2, \dots, n$, the denominator simplifies to the total number of observations $N$. Therefore, we have the closed-form solution:

\begin{align*}
\nu_k^{(t+1)} = \frac{\sum_{i=1}^n \gamma_{ik}^{(t)}}{n}, \quad k = 0, 1, \dots, K.
\end{align*}

\newpage

\bibliography{bibliography}

\begin{thebibliography}{}

\bibitem[Bauer et~al., 2020]{bauer2020simplifying}
Bauer, D.~J., Belzak, W.~C., and Cole, V.~T. (2020).
\newblock Simplifying the assessment of measurement invariance over multiple
  background variables: Using regularized moderated nonlinear factor analysis
  to detect differential item functioning.
\newblock {\em Structural Equation Modeling: a Multidisciplinary Journal},
  27(1):43--55.

\bibitem[Bechger and Maris, 2015]{bechger2015statistical}
Bechger, T.~M. and Maris, G. (2015).
\newblock A statistical test for differential item pair functioning.
\newblock {\em Psychometrika}, 80(2):317--340.

\bibitem[Belzak and Bauer, 2020]{belzak2020improving}
Belzak, W. and Bauer, D.~J. (2020).
\newblock Improving the assessment of measurement invariance: Using
  regularization to select anchor items and identify differential item
  functioning.
\newblock {\em Psychological Methods}, 25(6):673–690.

\bibitem[Bennink et~al., 2014]{bennink2014measuring}
Bennink, M., Croon, M.~A., Keuning, J., and Vermunt, J.~K. (2014).
\newblock Measuring student ability, classifying schools, and detecting item
  bias at school level, based on student-level dichotomous items.
\newblock {\em Journal of Educational and Behavioral Statistics},
  39(3):180--202.

\bibitem[Bhattacharya and McNicholas, 2014]{bhattacharya2014lasso}
Bhattacharya, S. and McNicholas, P.~D. (2014).
\newblock A lasso-penalized bic for mixture model selection.
\newblock {\em Advances in Data Analysis and Classification}, 8(1):45--61.

\bibitem[Birnbaum, 1968]{birnbaum1968some}
Birnbaum, A. (1968).
\newblock {\em Some latent trait models and their use in inferring an
  examinee's ability}, pages 397--472.
\newblock Addison-Wesley, Reading, MA.

\bibitem[Bock and Aitkin, 1981]{bock1981marginal}
Bock, R.~D. and Aitkin, M. (1981).
\newblock Marginal maximum likelihood estimation of item parameters:
  Application of an {EM} algorithm.
\newblock {\em Psychometrika}, 46(4):443--459.

\bibitem[Bolt et~al., 2002]{bolt2002item}
Bolt, D.~M., Cohen, A.~S., and Wollack, J.~A. (2002).
\newblock Item parameter estimation under conditions of test speededness:
  Application of a mixture {R}asch model with ordinal constraints.
\newblock {\em Journal of Educational Measurement}, 39(4):331--348.

\bibitem[Bouveyron and Brunet-Saumard, 2014]{bouveyron2014model}
Bouveyron, C. and Brunet-Saumard, C. (2014).
\newblock Model-based clustering of high-dimensional data: A review.
\newblock {\em Computational Statistics \& Data Analysis}, 71:52--78.

\bibitem[Candell and Drasgow, 1988]{candell1988iterative}
Candell, G.~L. and Drasgow, F. (1988).
\newblock An iterative procedure for linking metrics and assessing item bias in
  item response theory.
\newblock {\em Applied Psychological Measurement}, 12(3):253--260.

\bibitem[Cao et~al., 2017]{cao2017monte}
Cao, M., Tay, L., and Liu, Y. (2017).
\newblock A monte carlo study of an iterative wald test procedure for dif
  analysis.
\newblock {\em Educational and Psychological Measurement}, 77(1):104--118.

\bibitem[Chen et~al., 2023]{chen2021dif}
Chen, Y., Li, C., Ouyang, J., and Xu, G. (2023).
\newblock {DIF} statistical inference and detection without knowing anchoring
  items.
\newblock {\em Psychometrika}.
\newblock To appear.

\bibitem[Chen et~al., 2022]{chen2022detection}
Chen, Y., Lu, Y., and Moustaki, I. (2022).
\newblock Detection of two-way outliers in multivariate data and application to
  cheating detection in educational tests.
\newblock {\em The Annals of Applied Statistics}, 16(3):1718--1746.

\bibitem[Cho and Cohen, 2010]{cho2010multilevel}
Cho, S.-J. and Cohen, A.~S. (2010).
\newblock A multilevel mixture {IRT} model with an application to {DIF}.
\newblock {\em Journal of Educational and Behavioral Statistics},
  35(3):336--370.

\bibitem[Cho et~al., 2016]{cho2016ncme}
Cho, S.-J., Suh, Y., and Lee, W.-y. (2016).
\newblock An {NCME} instructional module on latent {DIF} analysis using mixture
  item response models.
\newblock {\em Educational Measurement: Issues and Practice}, 35(1):48--61.

\bibitem[Cizek and Wollack, 2017]{cizek2017handbook}
Cizek, G.~J. and Wollack, J.~A. (2017).
\newblock {\em Handbook of quantitative methods for detecting cheating on
  tests}.
\newblock Routledge New York, NY.

\bibitem[Clauser et~al., 1993]{clauser1993effects}
Clauser, B., Mazor, K., and Hambleton, R.~K. (1993).
\newblock The effects of purification of matching criterion on the
  identification of {DIF} using the {M}antel-{H}aenszel procedure.
\newblock {\em Applied Measurement in Education}, 6(4):269--279.

\bibitem[Cohen and Bolt, 2005]{cohen2005mixture}
Cohen, A.~S. and Bolt, D.~M. (2005).
\newblock A mixture model analysis of differential item functioning.
\newblock {\em Journal of Educational Measurement}, 42(2):133--148.

\bibitem[De~Boeck et~al., 2011]{de2011explanatory}
De~Boeck, P., Cho, S.-J., and Wilson, M. (2011).
\newblock Explanatory secondary dimension modeling of latent differential item
  functioning.
\newblock {\em Applied Psychological Measurement}, 35(8):583--603.

\bibitem[Dempster et~al., 1977]{dempster1977maximum}
Dempster, A.~P., Laird, N.~M., and Rubin, D.~B. (1977).
\newblock Maximum likelihood from incomplete data via the {EM} algorithm.
\newblock {\em Journal of the Royal Statistical Society: Series B
  (Methodological)}, 39(1):1--22.

\bibitem[Dorans and Kulick, 1986]{dorans1986demonstrating}
Dorans, N.~J. and Kulick, E. (1986).
\newblock Demonstrating the utility of the standardization approach to
  assessing unexpected differential item performance on the scholastic aptitude
  test.
\newblock {\em Journal of Educational Measurement}, 23(4):355--368.

\bibitem[Drabinov{\'a} and Martinkov{\'a}, 2017]{drabinova2017detection}
Drabinov{\'a}, A. and Martinkov{\'a}, P. (2017).
\newblock Detection of differential item functioning with nonlinear regression:
  A {N}on-{IRT} approach accounting for guessing.
\newblock {\em Journal of Educational Measurement}, 54(4):498--517.

\bibitem[Fan and Li, 2001]{fan2001variable}
Fan, J. and Li, R. (2001).
\newblock Variable selection via nonconcave penalized likelihood and its oracle
  properties.
\newblock {\em Journal of the American statistical Association},
  96(456):1348--1360.

\bibitem[Fidalgo et~al., 2000]{fidalgo2000effects}
Fidalgo, A., Mellenbergh, G.~J., and Mu{\~n}iz, J. (2000).
\newblock Effects of amount of {DIF}, test length, and purification type on
  robustness and power of {M}antel-{H}aenszel procedures.
\newblock {\em Methods of Psychological Research Online}, 5(3):43--53.

\bibitem[Finch and Hern{\'a}ndez~Finch, 2013]{finch2013investigation}
Finch, W.~H. and Hern{\'a}ndez~Finch, M.~E. (2013).
\newblock Investigation of specific learning disability and testing
  accommodations based differential item functioning using a multilevel
  multidimensional mixture item response theory model.
\newblock {\em Educational and Psychological Measurement}, 73(6):973--993.

\bibitem[Hastie et~al., 2009]{hastie2009elements}
Hastie, T., Tibshirani, R., Friedman, J.~H., and Friedman, J.~H. (2009).
\newblock {\em The Elements of Statistical Learning: Data Mining, Inference,
  and Prediction}.
\newblock Springer.

\bibitem[Holland and Thayer, 1986]{holland1986differential}
Holland, P.~W. and Thayer, D.~T. (1986).
\newblock Differential item functioning and the {M}antel-{H}aenszel procedure.
\newblock {\em ETS Research Report Series}, 1986(2):i--24.

\bibitem[Holland and Wainer, 1993]{holland1993differential}
Holland, P.~W. and Wainer, H. (1993).
\newblock {\em Differential Item Functioning}.
\newblock Psychology Press.

\bibitem[Hough et~al., 2001]{hough2001determinants}
Hough, L.~M., Oswald, F.~L., and Ployhart, R.~E. (2001).
\newblock Determinants, detection and amelioration of adverse impact in
  personnel selection procedures: Issues, evidence and lessons learned.
\newblock {\em International Journal of Selection and Assessment},
  9(1-2):152--194.

\bibitem[J{\"o}reskog and Goldberger, 1975]{joreskog1975estimation}
J{\"o}reskog, K.~G. and Goldberger, A.~S. (1975).
\newblock Estimation of a model with multiple indicators and multiple causes of
  a single latent variable.
\newblock {\em Journal of the American statistical Association},
  70(351a):631--639.

\bibitem[Kim et~al., 1995]{kim1995detection}
Kim, S.-H., Cohen, A.~S., and Park, T.-H. (1995).
\newblock Detection of differential item functioning in multiple groups.
\newblock {\em Journal of Educational Measurement}, 32(3):261--276.

\bibitem[Kopf et~al., 2015a]{kopf2015anchor}
Kopf, J., Zeileis, A., and Strobl, C. (2015a).
\newblock Anchor selection strategies for {DIF} analysis: Review, assessment,
  and new approaches.
\newblock {\em Educational and Psychological Measurement}, 75(1):22--56.

\bibitem[Kopf et~al., 2015b]{kopf2015framework}
Kopf, J., Zeileis, A., and Strobl, C. (2015b).
\newblock A framework for anchor methods and an iterative forward approach for
  {DIF} detection.
\newblock {\em Applied Psychological Measurement}, 39(2):83--103.

\bibitem[Kuha and Moustaki, 2015]{kuha2015nonequivalence}
Kuha, J. and Moustaki, I. (2015).
\newblock Nonequivalence of measurement in latent variable modeling of
  multigroup data: A sensitivity analysis.
\newblock {\em Psychological Methods}, 20(4):523--536.

\bibitem[Lord, 1977]{lord1977study}
Lord, F.~M. (1977).
\newblock A study of item bias, using item characteristic curve theory.
\newblock In Poortinga, Y.~H., editor, {\em Basic problems in cross-cultural
  psychology}, pages 19--29, Amsterdam. Swets \& Zeitlinger Publishers.

\bibitem[Lord, 1980]{lord1980applications}
Lord, F.~M. (1980).
\newblock {\em Applications of Item Response Theory to Practical Testing
  Problems}.
\newblock Routledge.

\bibitem[Luo et~al., 2008]{luo2008mixture}
Luo, R., Tsai, C.-L., and Wang, H. (2008).
\newblock On mixture regression shrinkage and selection via the mr-lasso.
\newblock {\em International Journal of Pure and Applied Mathematics},
  46:403--414.

\bibitem[Magis et~al., 2015]{magis2015detection}
Magis, D., Tuerlinckx, F., and De~Boeck, P. (2015).
\newblock Detection of differential item functioning using the lasso approach.
\newblock {\em Journal of Educational and Behavioral Statistics},
  40(2):111--135.

\bibitem[Millsap, 2012]{millsap2012statistical}
Millsap, R.~E. (2012).
\newblock {\em Statistical approaches to measurement invariance}.
\newblock Routledge.

\bibitem[Muthen and Lehman, 1985]{muthen1985multiple}
Muthen, B. and Lehman, J. (1985).
\newblock Multiple group {IRT} modeling: Applications to item bias analysis.
\newblock {\em Journal of Educational Statistics}, 10(2):133--142.

\bibitem[Muth{\'e}n, 1989]{muthen1989latent}
Muth{\'e}n, B.~O. (1989).
\newblock Latent variable modeling in heterogeneous populations.
\newblock {\em Psychometrika}, 54(4):557--585.

\bibitem[O’Leary and Smith, 2016]{o2016detecting}
O’Leary, L.~S. and Smith, R.~W. (2016).
\newblock Detecting candidate preknowledge and compromised content using
  differential person and item functioning.
\newblock In {\em Handbook of quantitative methods for detecting cheating on
  tests}, pages 151--163. Routledge.

\bibitem[Parikh and Boyd, 2014]{parikh2014proximal}
Parikh, N. and Boyd, S. (2014).
\newblock Proximal algorithms.
\newblock {\em Foundations and Trends{\textregistered} in Optimization},
  1(3):127--239.

\bibitem[Ployhart and Holtz, 2008]{ployhart2008diversity}
Ployhart, R.~E. and Holtz, B.~C. (2008).
\newblock The diversity--validity dilemma: Strategies for reducing racioethnic
  and sex subgroup differences and adverse impact in selection.
\newblock {\em Personnel Psychology}, 61(1):153--172.

\bibitem[Rasch, 1960]{rasch1960studies}
Rasch, G. (1960).
\newblock {\em Studies in mathematical psychology: I. Probabilistic models for
  some intelligence and attainment tests.}
\newblock Nielsen \& Lydiche.

\bibitem[Redner and Walker, 1984]{redner1984mixture}
Redner, R.~A. and Walker, H.~F. (1984).
\newblock Mixture densities, maximum likelihood and the {EM} algorithm.
\newblock {\em SIAM review}, 26(2):195--239.

\bibitem[Reeve and Teresi, 2016]{reeve2016overview}
Reeve, B.~B. and Teresi, J.~A. (2016).
\newblock Overview to the two-part series: Measurement equivalence of the
  {P}atient {R}eported {O}utcomes {M}easurement {I}nformation
  {S}ystem{\textregistered}({PROMIS}{\textregistered}) short forms.
\newblock {\em Psychological Test and Assessment Modeling}, 58(1):31--35.

\bibitem[Robitzsch, 2022]{robitzsch2022regularized}
Robitzsch, A. (2022).
\newblock Regularized mixture {R}asch model.
\newblock {\em Information}, 13(11):534.

\bibitem[Samejima, 1969]{samejima1969estimation}
Samejima, F. (1969).
\newblock Estimation of latent ability using a response pattern of graded
  scores.
\newblock {\em Psychometrika monograph supplement}.

\bibitem[Schauberger and Mair, 2020]{schauberger2020regularization}
Schauberger, G. and Mair, P. (2020).
\newblock A regularization approach for the detection of differential item
  functioning in generalized partial credit models.
\newblock {\em Behavior Research Methods}, 52(1):279--294.

\bibitem[Schwarz, 1978]{schwarz1978estimating}
Schwarz, G. (1978).
\newblock Estimating the dimension of a model.
\newblock {\em The Annals of Statistics}, 6(2):461--464.

\bibitem[Shao, 1997]{shao1997asymptotic}
Shao, J. (1997).
\newblock An asymptotic theory for linear model selection.
\newblock {\em Statistica Sinica}, 7(2):221--242.

\bibitem[Shealy and Stout, 1993]{shealy1993model}
Shealy, R. and Stout, W. (1993).
\newblock A model-based standardization approach that separates true bias/{DIF}
  from group ability differences and detects test bias/{DTF} as well as item
  bias/{DIF}.
\newblock {\em Psychometrika}, 58(2):159--194.

\bibitem[Steenkamp and Baumgartner, 1998]{steenkamp1998assessing}
Steenkamp, J.-B.~E. and Baumgartner, H. (1998).
\newblock Assessing measurement invariance in cross-national consumer research.
\newblock {\em Journal of Consumer Research}, 25(1):78--90.

\bibitem[Stephens, 2000]{stephens2000dealing}
Stephens, M. (2000).
\newblock Dealing with label switching in mixture models.
\newblock {\em Journal of the Royal Statistical Society: Series B (Statistical
  Methodology)}, 62(4):795--809.

\bibitem[Swaminathan and Rogers, 1990]{swaminathan1990detecting}
Swaminathan, H. and Rogers, H.~J. (1990).
\newblock Detecting differential item functioning using logistic regression
  procedures.
\newblock {\em Journal of Educational Measurement}, 27(4):361--370.

\bibitem[Tay et~al., 2016]{tay2016item}
Tay, L., Huang, Q., and Vermunt, J.~K. (2016).
\newblock Item response theory with covariates ({IRT-C}) assessing item
  recovery and differential item functioning for the three-parameter logistic
  model.
\newblock {\em Educational and Psychological Measurement}, 76(1):22--42.

\bibitem[Tay et~al., 2015]{tay2015overview}
Tay, L., Meade, A.~W., and Cao, M. (2015).
\newblock An overview and practical guide to irt measurement equivalence
  analysis.
\newblock {\em Organizational Research Methods}, 18(1):3--46.

\bibitem[Teresi and Reeve, 2016]{teresi2016epilogue}
Teresi, J.~A. and Reeve, B.~B. (2016).
\newblock Epilogue to the two-part series: Measurement equivalence of the
  {P}atient {R}eported {O}utcomes {M}easurement {I}nformation
  {S}ystem{\textregistered}({PROMIS}{\textregistered}) short forms.
\newblock {\em Psychological Test and Assessment Modeling}, 58(2):423--433.

\bibitem[Teresi et~al., 2021]{teresi2021differential}
Teresi, J.~A., Wang, C., Kleinman, M., Jones, R.~N., and Weiss, D.~J. (2021).
\newblock Differential item functioning analyses of the {P}atient-{R}eported
  {O}utcomes {M}easurement {I}nformation {S}ystem ({PROMIS}{\textregistered})
  measures: methods, challenges, advances, and future directions.
\newblock {\em Psychometrika}, 86(3):674--711.

\bibitem[Thissen and Steinberg, 1988]{thissen1988data}
Thissen, D. and Steinberg, L. (1988).
\newblock Data analysis using item response theory.
\newblock {\em Psychological Bulletin}, 104(3):385--395.

\bibitem[Thissen et~al., 1988]{thissen2013use}
Thissen, D., Steinberg, L., and Wainer, H. (1988).
\newblock Use of item response theory in the study of group differences in
  trace lines.
\newblock In Wainer, H. and Braun, H.~I., editors, {\em Test Validity}, pages
  147--172, Hillsdale, NJ. Lawrence Erlbaum Associates, Inc.

\bibitem[Tibshirani, 1996]{tibshirani1996regression}
Tibshirani, R. (1996).
\newblock Regression shrinkage and selection via the lasso.
\newblock {\em Journal of the Royal Statistical Society: Series B
  (Methodological)}, 58(1):267--288.

\bibitem[Tutz and Schauberger, 2015]{tutz2015penalty}
Tutz, G. and Schauberger, G. (2015).
\newblock A penalty approach to differential item functioning in {R}asch
  models.
\newblock {\em Psychometrika}, 80(1):21--43.

\bibitem[van~de Geer, 2008]{van2008high}
van~de Geer, S.~A. (2008).
\newblock High-dimensional generalized linear models and the lasso.
\newblock {\em The Annals of Statistics}, 36(2):614--645.

\bibitem[Vermunt, 2010]{vermunt2010latent}
Vermunt, J.~K. (2010).
\newblock Latent class modeling with covariates: Two improved three-step
  approaches.
\newblock {\em Political Analysis}, 18(4):450--469.

\bibitem[Vermunt and Magidson, 2021]{vermunt2021perform}
Vermunt, J.~K. and Magidson, J. (2021).
\newblock How to perform three-step latent class analysis in the presence of
  measurement non-invariance or differential item functioning.
\newblock {\em Structural Equation Modeling: A Multidisciplinary Journal},
  28(3):356--364.

\bibitem[Von~Davier et~al., 2011]{von2011measuring}
Von~Davier, M., Xu, X., and Carstensen, C.~H. (2011).
\newblock Measuring growth in a longitudinal large-scale assessment with a
  general latent variable model.
\newblock {\em Psychometrika}, 76(2):318--336.

\bibitem[Wainer, 2012]{wainer2012item}
Wainer, H. (2012).
\newblock An item response theory model for test bias and differential test
  functioning.
\newblock In {\em Differential Item Functioning}, pages 202--244. Routledge.

\bibitem[Wang et~al., 2023]{wang2021using}
Wang, C., Zhu, R., and Xu, G. (2023).
\newblock Using lasso and adaptive lasso to identify dif in multidimensional
  2pl models.
\newblock {\em Multivariate Behavioral Research}, 58(2):387--407.

\bibitem[Wang et~al., 2009]{wang2009mimic}
Wang, W.-C., Shih, C.-L., and Yang, C.-C. (2009).
\newblock The mimic method with scale purification for detecting differential
  item functioning.
\newblock {\em Educational and Psychological Measurement}, 69(5):713--731.

\bibitem[Wang and Su, 2004]{wang2004effects}
Wang, W.-C. and Su, Y.-H. (2004).
\newblock Effects of average signed area between two item characteristic curves
  and test purification procedures on the {DIF} detection via the
  {M}antel-{H}aenszel method.
\newblock {\em Applied Measurement in Education}, 17(2):113--144.

\bibitem[Wang and Yeh, 2003]{wang2003effects}
Wang, W.-C. and Yeh, Y.-L. (2003).
\newblock Effects of anchor item methods on differential item functioning
  detection with the likelihood ratio test.
\newblock {\em Applied Psychological Measurement}, 27(6):479--498.

\bibitem[Woods, 2009]{woods2009evaluation}
Woods, C.~M. (2009).
\newblock Evaluation of mimic-model methods for dif testing with comparison to
  two-group analysis.
\newblock {\em Multivariate Behavioral Research}, 44(1):1--27.

\bibitem[Woods et~al., 2013]{woods2013langer}
Woods, C.~M., Cai, L., and Wang, M. (2013).
\newblock The langer-improved wald test for {DIF} testing with multiple groups:
  Evaluation and comparison to two-group {IRT}.
\newblock {\em Educational and Psychological Measurement}, 73(3):532--547.

\bibitem[Woods and Grimm, 2011]{woods2011testing}
Woods, C.~M. and Grimm, K.~J. (2011).
\newblock Testing for nonuniform differential item functioning with multiple
  indicator multiple cause models.
\newblock {\em Applied Psychological Measurement}, 35(5):339--361.

\bibitem[Yuan et~al., 2021]{yuan2021differential}
Yuan, K.-H., Liu, H., and Han, Y. (2021).
\newblock Differential item functioning analysis without a priori information
  on anchor items: {QQ} plots and graphical test.
\newblock {\em Psychometrika}, 86(2):345--377.

\bibitem[Zhang, 2010]{zhang2010nearly}
Zhang, C.-H. (2010).
\newblock Nearly unbiased variable selection under minimax concave penalty.
\newblock {\em The Annals of Statistics}, 38(2):894--942.

\bibitem[Zhao and Yu, 2006]{zhao2006model}
Zhao, P. and Yu, B. (2006).
\newblock On model selection consistency of {L}asso.
\newblock {\em The Journal of Machine Learning Research}, 7:2541--2563.

\bibitem[Zhao et~al., 2021]{zhao2021defense}
Zhao, S., Witten, D., and Shojaie, A. (2021).
\newblock In defense of the indefensible: A very naive approach to
  high-dimensional inference.
\newblock {\em Statistical Science}, 36(4):562--577.

\bibitem[Zou, 2006]{zou2006adaptive}
Zou, H. (2006).
\newblock The adaptive lasso and its oracle properties.
\newblock {\em Journal of the American Statistical Association},
  101(476):1418--1429.

\bibitem[Zwick et~al., 2000]{zwick2000using}
Zwick, R., Thayer, D.~T., and Lewis, C. (2000).
\newblock Using loss functions for {DIF} detection: An empirical {B}ayes
  approach.
\newblock {\em Journal of Educational and Behavioral Statistics},
  25(2):225--247.

\end{thebibliography}

\end{document}